\documentclass[12pt]{article}
\usepackage{amssymb,graphicx,color,amsfonts,amsmath}
\begin{document}

\baselineskip=.22in
\renewcommand{\baselinestretch}{1.2}
\renewcommand{\theequation}{\thesection.\arabic{equation}}
~\vspace{6mm}

\begin{center}
{{{\Large \bf BPS Vortices, $Q$-balls, and $Q$-vortices
\\[2mm]{\Large \bf in ${\cal N}=6$} Chern-Simons Matter Theory}
}\\[12mm]
{Gyungchoon Go${}^{1}$, Chanju Kim${}^{2}$, Yoonbai Kim${}^{3}$,
O-Kab Kwon${}^{3}$, Hiroaki Nakajima${}^{4}$}\\[5mm]
{\it ${}^{1}$Center for Nanotubes and Nanostructured Composites,\\
Sungkyunkwan University, Suwon 440-746, Korea}\\
{\tt gcgo@skku.edu}\\[3mm]
{\it ${}^{2}$Institute for the Early Universe and Department of Physics,\\
Ewha Womans University, Seoul 120-750, Korea}\\
{\tt cjkim@ewha.ac.kr}\\[3mm]
{\it ${}^{3}$Department of Physics, BK21 Physics Research Division,
and Institute of Basic Science\\
Sungkyunkwan University, Suwon 440-746, Korea}\\
{\tt yoonbai, okab@skku.edu}\\[3mm]
{\it ${}^{4}$Department of Physics
and Center for Theoretical Sciences,\\
National Taiwan University,
Taipei 10617, Taiwan, R.O.C.}\\
{\tt nakajima@phys.ntu.edu.tw} }
\end{center}
\vspace{10mm}

\begin{abstract}
We investigate the vortex-type BPS equations in the ABJM theory without and with mass-deformation.
We systematically classify the BPS equations in terms of the number of supersymmetry
and the R-symmetries of the undeformed and mass-deformed ABJM theories.
For the undeformed case, we analyze the ${\cal N}=2$ BPS equations for U(2)$\times$U(2)
gauge symmetry and obtain a coupled differential equation which can be reduced
to either Liouville- or Sinh-Gordon-type vortex equations according to the choice of
scalar functions.
For the mass-deformed case with U($N$)$\times$U($N$) gauge symmetry, we obtain some number of pairs of coupled differential equations from the ${\cal N}=1,2$ BPS equations,
which can be reduced to the vortex equations in Maxwell-Higgs
theory or Chern-Simons matter theories as special cases.
We discuss the solutions.
In ${\cal N}=3$ vortex equations Chern-Simons-type vortex equation is not allowed.
We also show that ${\cal N}=\frac52, \frac32, \frac12$ BPS equations
are equivalent to those with higher integer supersymmetries.

\end{abstract}


\newpage

\tableofcontents

\setcounter{equation}{0}
\section{Introduction}\label{sec1}

After the first construction of the ${\cal N}=8$ superconformal
Chern-Simons theory (SCS) based on the three algebra
by Bagger-Lambert-Gustavsson (BLG)~\cite{Bagger:2006sk,Gustavsson:2007vu},
an ${\cal N}=6$ SCS theory with U($N$)$\times$U($N$)
gauge symmetry was constructed by Aharony-Bergman-Jafferis-Maldacena (ABJM)~\cite{Aharony:2008ug}.
The latter theory includes a large class of SCS theories depending
on the rank of the gauge group $N$ and the Chern-Simons level $k$.
This ABJM theory was proposed as a low energy effective action of $N$ coincident M2-branes
on the ${\rm {\mathbb C}}^4/{\mathbb Z}_k$ orbifold fixed point and was conjectured to be dual
to type IIA string theory on ${\rm AdS}_4\times \mathbb{CP}^3$ ($N^{\frac15}\ll k\ll N$)
or M-theory on ${\rm AdS}_4\times {\rm S}^7/\mathbb{Z}_k$ $(k\ll N^{\frac15})$
in the large $N$ limit.

There has been significant progress in understanding the dynamics of M2-branes in M-theory
by the BLG and ABJM theories.
One direction of this progress might be obtaining solitonic objects which can be identified
with M-theory branes, such as M2- and M5-branes.
In the BLG theory, the composite of M2- and M5-branes~\cite{Krishnan:2008zm,Bonelli:2008kh},
the domain wall solutions~\cite{Hosomichi:2008qk},
and some vortex-type BPS configurations~\cite{Kim:2008cp} were obtained without and with mass-deformation
and also possible BPS equations were classified~\cite{Jeon:2008bx,Jeon:2008zj}.
Similarly, in the ABJM theory, the composite of M-branes and domain wall
solutions~\cite{Terashima:2008sy,Hanaki:2008cu,Mohammed:2010eb},
the vortex-type solutions~\cite{Arai:2008kv,Kim:2009ny,Auzzi:2009es},
and the classification of BPS conditions of
intersecting M-branes~\cite{Fujimori:2010ec} were studied.
For the vortex-type solutions in the ABJM theory,
${\cal N}=1$ Chern-Simons vortex-type BPS equations~\cite{Arai:2008kv}
and Yang-Mills Higgs vortex-type ${\cal N}=3$ BPS
ones were obtained~\cite{Kim:2009ny,Auzzi:2009es} and the existence
of the corresponding vortices with ${\cal N}=3$ supersymmetry was discussed~\cite{Han:2012wh}.
Vortex solutions in the nonrelativistic limit of the ABJM theory have been studied in
Ref.~\cite{Kawai:2009rc}.

As we already know, the mass-deformed BLG and ABJM theories have sextic bosonic potentials with
the symmetric and broken vacua.
In this reason, these theories can be understood as more complicated Chern-Simons-Higgs theories.
So in specific limits of the vortex-type BPS equations in the BLG and ABJM theory, one can obtain
the vortex configurations which were widely studied in Abelian and non-Abelian Chern-Simon-Higgs
theories~\cite{Hong:1990yh,Lee:1990it,Jackiw:1990pr,Kim:1992uw,Cugliandolo:1990nb}.

In Ref.~\cite{Fujimori:2010ec} the BPS equations preserving
various supersymmetries in the undeformed ABJM theory were
classified and the corresponding BPS configurations were interpreted
as known BPS objects in M-theory.
In this paper, we recapitulate the classification of the vortex-type BPS
configurations of the undeformed and mass-deformed ABJM theories
in terms of the number of remaining supersymmetry and
SU(4) and SU(2)$\times$SU(2)$\times$U(1) R-symmetries, respectively.
Then we reproduce the resulting BPS equations by reshuffling the energy
expression, thereby obtaining the BPS energy bound. An unusual quantity
we consider here is the stress tensor which is the spatial component
of the energy-momentum tensor. It has been argued \cite{Moreno:2008me}
from the viewpoint of supersymmetry algebra that the stress vanishes
for BPS configurations. We explicitly check it for the ABJM theory.
We will see that this can actually be a useful method to obtain consistency
conditions for BPS equations in case of lower supersymmetries.

The main concern of the paper is to study possible solutions of the
vortex-type BPS equations for various supersymmetries.
We have already considered the half-BPS ($\mathcal{N}=3$) case in
\cite{Kim:2009ny}. In this paper we extend our analysis to less supersymmetric
cases: $\mathcal{N}=2,1$ and $\mathcal{N}=5/2,3/2,1/2$.
Since the BPS equations get complicated as the number of supersymmetries
are smaller, it would not be feasible to find the most general solutions.
With suitable ansatzes, however, we show that for some $\mathcal{N}=2$ and
$\mathcal{N}=1$ cases the equations reduce to well-known equations
such as the vortex equation in the Maxwell-Higgs theories and
Chern-Simons-Higgs theories \cite{Hong:1990yh,Jackiw:1990pr,Kim:1992uw}.
We compute the energies and the angular momenta for the solutions.

For half-integer supersymmetric cases, we show that the supersymmetries
of the solutions to the BPS equations in mass-deformed theory are actually
enhanced to integer ones. For example, $\mathcal{N}=5/2$ BPS equations
are shown to be identical to $\mathcal{N}=3$ BPS equations. Also
$\mathcal{N}=3/2$ cases are enhanced to either $\mathcal{N}=3$ or
$\mathcal{N}=2$ cases depending on the supersymmetric conditions. Similarly
$\mathcal{N}=1/2$ BPS equations are equivalent to $\mathcal{N}=1$ BPS
equations. It turns out that the stress tensor is a useful quantity
to show these enhancement.

The remaining part of this paper is organized as follows.
In section 2 we briefly review the ABJM theory without and with
mass-deformation to fix the notation.
In section 3 we recapitulate the ${\cal N}=3$ vortex-type BPS equations
obtained in Refs.~\cite{Kim:2009ny,Auzzi:2009es} and consider more general
BPS configurations with finite energy bounds in the mass-deformed case.
In sections 4 and 5 we systematically analyze the ${\cal N}=2$ and
${\cal N}=1$ BPS equations, respectively.
For the undeformed case in section 4, we obtain a coupled differential
equation which can be reduced to either Liouville- or Sinh-Gordon-type
vortex equations in special choices of scalar functions.
For the mass-deformed case, we obtain some number
of pairs of coupled differential equations which have appeared in
U($1$)$\times$U($1$) Chern-Simons system \cite{Kim:1993mh}.
We show that these differential equations are reduced to vortex equations
in Maxwell-Higgs theory or Chern-Simons matter theories,
according to the choices of scalar functions.
Difference between ${\cal N}=2$ case and ${\cal N}=1$ case is also discussed.
In section 6 we show that the ${\cal N}=\frac52, \frac32,\frac12$ BPS
equations are equivalent to those with higher integer supersymmetries.
Section 7 is devoted to conclusions and discussions.
In appendix A we add the analyses of two more cases of BPS vortex equations with ${\cal N}=1,2$ symmetries.

\setcounter{equation}{0}
\section{ABJM Theory without and with Mass Deformation}\label{sec2}
In this section we briefly review the ABJM theory and its mass deformation
to fix the notation.
The ABJM theory is an ${\cal N}=6$ superconformal ${\rm U}(N)\times
{\rm U}(N)$ Chern-Simons gauge theory with level $(k,-k)$, coupled
to four complex scalars and four fermions in the bifundamental
representation,
\begin{align}\label{ABJMa}
S_{{\rm ABJM}} =  \int d^3x\: \bigg\{
    &\frac{k}{4\pi} \epsilon^{\mu\nu\lambda} {\rm tr}\left(
           A_\mu \partial_\nu A_\lambda + \frac{2i}{3} A_\mu A_\nu A_\lambda
         - \hat{A}_\mu \partial_\nu \hat{A}_\lambda - \frac{2i}{3}
         \hat{A}_\mu \hat{A}_\nu \hat{A}_\lambda\right)
\nonumber \\
& - {\rm tr} \big(D_\mu Y_{A}^\dagger D^\mu Y^{A} \big)
    + {\rm tr}\big( \psi^{A\dagger} i\gamma^{\mu}D_{\mu} \psi_{A}\big)
    - V_{{\rm ferm}} - V_0 \bigg\},
\end{align}
where $A = 1,\ldots,4$ and
\begin{equation}
D_{\mu}Y^A = \partial_{\mu} Y^A + iA_{\mu}Y^A-iY^A{\hat A}_{\mu}.\nonumber
\end{equation}
We choose real gamma matrices $\gamma^{\mu}$ with the convention
$\gamma^2 = \gamma^0 \gamma^1$. An explicit representation would
be
\begin{align}
\gamma^{0}=i\sigma^{2}, \quad \gamma^{1}=\sigma^{1}, \quad
\gamma^{2}=\sigma^{3}.
\end{align}
The products of spinors are expressed by $\xi \chi\equiv
\xi^\alpha\chi_\alpha$ and
$\xi\gamma^\mu\chi=\xi^\alpha\gamma_{\alpha}^{\mu\,\beta}\chi_\beta$
with explicit spinor indices for two component spinors $\xi$ and
$\chi$, $\alpha$, $\beta\,$=$\,1$,$\,2$.

In the action \eqref{ABJMa}, $V_{{\rm ferm}}$ is the Yukawa-type
quartic-interaction term,
\begin{align}
V_{{\rm ferm}}=\frac{2i\pi}{k}\, {\rm tr}\big(
&Y_{A}^{\dagger}Y^{A}\psi^{B\dagger}\psi_{B}
-Y^{A}Y_{A}^{\dagger}\psi_{B}\psi^{B\dagger}
+2Y^{A}Y_{B}^{\dagger}\psi_{A}\psi^{B\dagger}
-2Y_{A}^{\dagger}Y^{B}\psi^{A\dagger}\psi_{B} \nonumber\\
& -\epsilon^{ABCD}Y_{A}^{\dagger}\psi_{B}Y_{C}^{\dagger}\psi_{D}
+\epsilon_{ABCD}Y^{A}\psi^{B\dagger}Y^{C}\psi^{D\dagger}
\big),
\end{align}
and $V_0$ is the sixth-order scalar potential,
\begin{align} \label{potential0}
V_0= -\frac{4\pi^{2}}{3k^{2}}\, {\rm tr}\big(&
Y^{A}Y_{A}^{\dagger}Y^{B}Y_{B}^{\dagger}Y^{C}Y_{C}^{\dagger}
+Y_{A}^{\dagger}Y^{A}Y_{B}^{\dagger}Y^{B}Y_{C}^{\dagger}Y^{C} \nonumber\\
& +4Y^{A}Y_{B}^{\dagger}Y^{C}Y_{A}^{\dagger}Y^{B}Y_{C}^{\dagger}
-6Y^{A}Y^{\dagger}_{B}Y^{B}Y_{A}^{\dagger}Y^{C}Y_{C}^{\dagger}
\big).
\end{align}
It can be written in a manifestly positive-definite form
\cite{Bandres:2008ry,Hosomichi:2008ip},
\begin{equation} \label{potential11}
V_0 = \frac23 {\rm tr} \left|\beta^{BC}_{\;A}
             + \delta^{[B}_A \beta^{C]D}_{\;D} \right|^2,
\end{equation}
where we have introduced the notation
$|\mathcal{O}|^2 \equiv {\cal O}^\dagger {\cal O}$, and $\beta_C^{AB}$ is
defined by
\begin{equation}
\beta^{AB}_{\; C}
  = \frac{4\pi}{k}Y^{[A}Y_{C}^{\dagger}Y^{B]}.
\label{bABC}
\end{equation}

By adding mass terms to the action, the theory can be
deformed~\cite{Hosomichi:2008jb,Gomis:2008vc} in the unique
way which preserve the full ${\cal N}=6$ supersymmetry~\cite{Hosomichi:2008qk},
\begin{align}
&\Delta V_{{\rm ferm}}= {\rm tr}\, \mu \psi^{\dagger A}M_{A}^{\;
B}\psi_{B},
\nonumber\\
&\Delta V_0={\rm tr}\left( \frac{4\pi\mu}{k}Y^{A}
Y^{\dagger}_{A}Y^{B}M_{B}^{\; C}Y_{C}^{\dagger}
-\frac{4\pi\mu}{k}Y^{\dagger}_{A}Y^{A}Y^{\dagger}_{B}M_{C}^{\;
B}Y^{C} +\mu^{2}Y^{\dagger}_{A}Y^{A} \right), \label{md2}
\end{align}
where $\mu$ is the mass deformation parameter and $M_A^B =
\rm{diag}(1,1,-1,-1)$. Combined with (\ref{potential11}),
the potential $V_\textrm{m}$ in the mass-deformed
theory can also be written in a manifestly positive-definite
form \cite{Kim:2009ny},
\begin{equation} \label{potential3}
V_\textrm{m} = V_0+\Delta V_0 = \frac23 {\rm tr} \left|\beta^{BC}_{\;A}
       + \delta^{[B}_A \beta^{C]D}_{\;D} + \mu M_A^{\;[B} Y^{C]} \right|^2.
\end{equation}
It is not difficult to see that the theory
is invariant under the following ${\cal N}=6$ supersymmetry
transformation~\cite{Aharony:2008ug,Hosomichi:2008qk,Gaiotto:2008cg,Terashima:2008sy},
\begin{align} \label{strf}
\delta Y^{A}&= i\omega^{AB}\psi_{B}, \nonumber\\
\delta \psi_{A}&= \gamma^{\mu}\omega_{AB}D_{\mu}Y^{B}+\frac{2\pi}{k}\left[
-\omega_{AB}\big(Y^{C}Y_{C}^{\dagger}Y^{B}-Y^{B}Y_{C}^{\dagger}Y^{C}\big)
+2\omega_{BC}Y^{B}Y_{A}^{\dagger}Y^{C}\right] + \mu M_A^{\;B} \omega_{BC}Y^{C}
     \nonumber \\
               &= \gamma^{\mu}\omega_{AB}D_{\mu}Y^{B}
    +\omega_{BC} \left( \beta^{BC}_{\; A}
                       + \delta_A^{[B} \beta^{C]D}_{\; D} \right)
    + \mu M_A^{\;B} \omega_{BC}Y^{C}, \nonumber\\
\delta A_\mu &= -\frac{2\pi}{k}\big(Y^{A}\psi^{B\dagger}\gamma_{\mu}
\omega_{AB}+\omega^{AB}\gamma_{\mu}\psi_{A}Y_{B}^{\dagger}\big), \nonumber \\
\delta \hat A_\mu &= \frac{2\pi}{k}\big(\psi^{A\dagger}Y^{B}\gamma_{\mu}
\omega_{AB}+\omega^{AB}\gamma_{\mu}Y_{A}^{\dagger}\psi_{B}\big),
\end{align}
where $\omega_{AB}$ are supersymmetry transformation parameters with
\begin{equation} \label{omegaab}
\omega^{AB} = \omega_{AB}^* = \frac12 \epsilon^{ABCD}\omega_{CD}.
\end{equation}
Note that the mass deformation affects only the transformation of the
fermion fields by an additional term,
\begin{equation} \label{delmpsi}
\delta_{{\rm m}}\psi_{A}=\mu M_A^{\;B} \omega_{BC}Y^{C}.
\end{equation}
Equation \eqref{md2} is not the only form of the mass-deformed theory.
One can also get mass-deformed theories in $\mathcal N =1$ or $\mathcal N =2$
superfield formalism for which only part of the supersymmetry is manifest.
It can, however, be shown \cite{Kim:2009ny} that they are all equivalent to
\eqref{md2} by a suitable field redefinition.

From \eqref{potential3} the vacuum equation of the mass-deformed theory is
\begin{equation}
\beta^{BC}_{\;A}
       + \delta^{[B}_A \beta^{C]D}_{\;D} + \mu M_A^{\;[B} Y^{C]}=0,
\end{equation}
which reduces to \cite{Gomis:2008vc,Kim:2009ny}
\begin{align}
&\beta^{ab}_{a}+\mu Y^b=0,\nonumber\\& \beta^{pq}_{p}-\mu
Y^q=0,\nonumber\\
&\beta^{ba}_{p}=\beta^{qa}_{p}=\beta^{pq}_{a}=\beta^{ab}_{p}=0.\label{ve3}
\end{align}
where $a,b=1,2$ and $p,q=3,4$.
The general solution of these vacuum equations was found in \cite{Gomis:2008vc} and refined in
\cite{Kim:2010mr},
\begin{align}\label{gvac}
Y^{a}&=\sqrt{ \frac{k \mu}{2\pi} }\left(\begin{array}{c}
\begin{array}{cccccc}{\bf 0}_{m_1\times (m_1+1)}&&&&&\\&\ddots&&&&\\
&&{\bf 0}_{m_I\times(m_I + 1)} &&&\\
&&& {\mathcal M^\dag}_a^{(m_{I+1})}\!\!&&\\&&&&\!\!\ddots\!&\\
&&&&&\!\!{\mathcal M^\dag}_a^{(m_f)}\end{array}\\
\end{array}\right),\nonumber \\
Y^{a+2}&=\sqrt{ \frac{k \mu}{2\pi} } \left(\begin{array}{c}
\begin{array}{cccccc}\mathcal{M}_a^{(m_1)}\!\!&&&&&\\&\!\!\ddots\!&&&&\\
&&\!\!\mathcal{M}_a^{(m_I)}&&& \\ &&& {\bf 0}_{(m_{I+1}+1)\times m_{I+1}}
&&\\&&&&\ddots&\\&&&&&{\bf 0}_{(m_f+1)\times m_f}\end{array}\\
\end{array}\right),
\end{align}
where ${\cal M}_{a}^{m}$ is an $m\times (m+1)$ matrix,
\begin{align}
{\cal M}_1^{m}=\sqrt{ \frac{k \mu}{2\pi} }\begin{pmatrix}
  \sqrt{m} &0 & & &\\&\hspace{-4mm}\sqrt{m-1} &0& &
  &\\& &\ddots  &\ddots &
  &\\& & &\sqrt{2}&0 &\\& & & &1 &0
  \end{pmatrix},~~{\cal M}_2^{m}=\sqrt{ \frac{k \mu}{2\pi} }\begin{pmatrix}
  0 &1 & & &\\&0 &\sqrt{2}& &
  &\\& &\ddots  &\ddots &
  &\\& & &\hspace{-5mm}0&\hspace{-5mm}\sqrt{m-1} &\\& & & &0 &\hspace{-5mm}\sqrt{m}
  \end{pmatrix}.
\end{align}
For the U($N)\times {\rm U}(N)$ gauge group we have the following constraints
\begin{align}
\sum_{m=0}^\infty\big[m \tilde N_m + (m+1) \hat N_m\big] = N,\quad
\sum_{m=0}^\infty\big[(m+1) \tilde N_m + m \hat N_m\big] = N,
\end{align}
where $\tilde N_m$ and $\hat N_m$ denote the numbers of block of ${\mathcal
M}_a^{(m)}$ and ${\mathcal M^\dag}_a^{(m)}$-types, and
$\tilde N_0$ and $\hat N_0$ represent the numbers of empty columns and empty rows, respectively.

Since we are interested in the classical vortex-type configurations,
we consider the Euler-Lagrange equations of gauge fields $A^\mu$ and $\hat A^\mu$
\begin{align}
\frac{k}{2\pi}\epsilon^{\mu\nu\lambda}F_{\nu\lambda}=j^\mu,\qquad \frac{k}{2\pi}\epsilon^{\mu\nu\lambda}{\hat
F}_{\nu\lambda}=-\hat j^\mu,\label{Aeq}
\end{align}
where
\begin{alignat}{3}
F_{\mu\nu}&=\partial_\mu A_\nu- \partial_\nu A_\mu+i[A_\mu,A_\nu],
& \hat F_{\mu\nu} &=\partial_\mu {\hat A}_\nu
            - \partial_\nu \hat A_\mu+i[\hat A_\mu,\hat A_\nu],\notag\\
j^\mu&=i\left[(D^{\mu}Y_A^\dag)Y^A-Y_A^\dag(D^\mu Y^A)\right],
& \hat j^\mu &=i\left[(D^{\mu}Y^A)Y_A^\dag-Y^A(D^\mu Y_A^\dag)\right]
\label{Ucu}.
\end{alignat}
The U(1) currents are obtained by taking trace,
\begin{equation}
j^\mu_{{\rm U}(1)}={\rm tr}\,j^\mu,\qquad
{\hat j}^\mu_{{\rm U}(1)}={\rm tr}\,{\hat j}^\mu,\label{jmu}
\end{equation}
and the corresponding charges are
\begin{equation}
Q=\int d^2 x \, j^0_{{\rm U}(1)},\qquad {\hat Q}=\int d^2 x\,  {\hat j}^0_{{\rm U}(1)}.\label{u1c}
\end{equation}

The Gauss' laws are the time components of \eqref{Aeq}
\begin{align}
B&=\frac{2\pi}{k}j^0=i\frac{2\pi}{k}\left[Y^A (D^{0}Y_A^\dag) - (D^0
Y^A) Y_A^\dag\right],\label{Beq} \\
\hat B&=-\frac{2\pi}{k}\hat j^0=-i\frac{2\pi}{k}\left[Y_A^\dag
(D^{0}Y^A) - (D^0 Y_A^\dag) Y^A\right],\label{Bheq}
\end{align}
where $B=\partial_1 A_2 -\partial_2 A_1+i[A_1,A_2]$
and $\hat B=\partial_1 \hat A_2 - \partial_2 \hat A_1+i[\hat A_1, \hat A_2]$
are magnetic fields. The spatial integral of the left-hand
sides of (\ref{Beq})--(\ref{Bheq}) gives non-abelian magnetic fluxes
\begin{align}
\Phi=\int d^2 x  B,
\qquad {\hat \Phi}=\int d^2 x  {\hat B},
\end{align}
and taking trace leads to U(1) magnetic fluxes,
\begin{align}
\Phi_{{\rm U}(1)}=&{\rm tr}\,\Phi
=\oint_{|x^{i}|\rightarrow\infty}\hspace{-3mm}dx^{i}
{\rm tr}\, A^{i},\qquad
{\hat \Phi}_{{\rm U}(1)}={\rm tr}\,{\hat \Phi}=\oint_{|x^{i}|\rightarrow\infty}\hspace{-3mm}
dx^{i}{\rm tr}\, {\hat A}^{i}.\label{mfu1}
\end{align}

As we shall see in the subsequent sections, the BPS equations can
also be derived by reshuffling the bosonic sector of energy-momentum
tensor
\begin{align}\label{eneT}
T^{\mu\nu}
={\rm tr}(D^{\mu}Y_A^\dag D^{\nu}Y^A+D^{\nu}Y_A^\dag
D^{\mu}Y^A)-\eta^{\mu\nu}\left[{\rm tr}(D^{\mu}Y_A^\dag
D_{\mu}Y^A+V_0)\right].
\end{align}
For later convenience we introduce energy, linear
momentum, and angular momentum, respectively
\begin{align}
&E = \int d^2x\, T^{00}=\int d^2x\left[{\rm tr}\,(D^{0}Y_A^\dag
D^{0}Y^A)+{\rm tr}\,(D^{i}Y_A^\dag
D^{i}Y^A)+V_m\right],\label{Ene}\\
&p^i=\int d^2x\, T^{i0}=\int d^2x\,{\rm tr}\,(D^{0}Y_A^\dag
D^{i}Y^A+D^{i}Y_A^\dag D^{0}Y^A),\label{limom}\\
&J=\int d^2x \,\epsilon_{ij}\, x^i T^{j0} =\int d^2x \,\epsilon_{ij}
x^i {\rm tr}\,(D^{j}Y_A^\dag D^{0}Y^A+D^{0}Y_A^\dag D^{j}Y^A).
\label{anm}
\end{align}
Pressure component is given by every spatial diagonal component of
the energy-momentum tensor and spatial stress is obtained from the
off-diagonal component,
\begin{equation}\label{stress}
P^i\equiv T^{ii}\,\,\,\,\,\,({\rm no \,\, sum \,\, over\,\,
}i),\qquad T^{ij}={\rm tr}(D^{i}Y_A^\dag D^{j}Y^A+D^j Y_A^\dag D^i
Y^A)\,\,\,\,\,\, (i\neq j).
\end{equation}

\begin{figure}[h]\centering
\vspace{0mm} \scalebox{1.2}[1.2]{
\scalebox{0.7}[0.7]{\includegraphics[width=250mm]{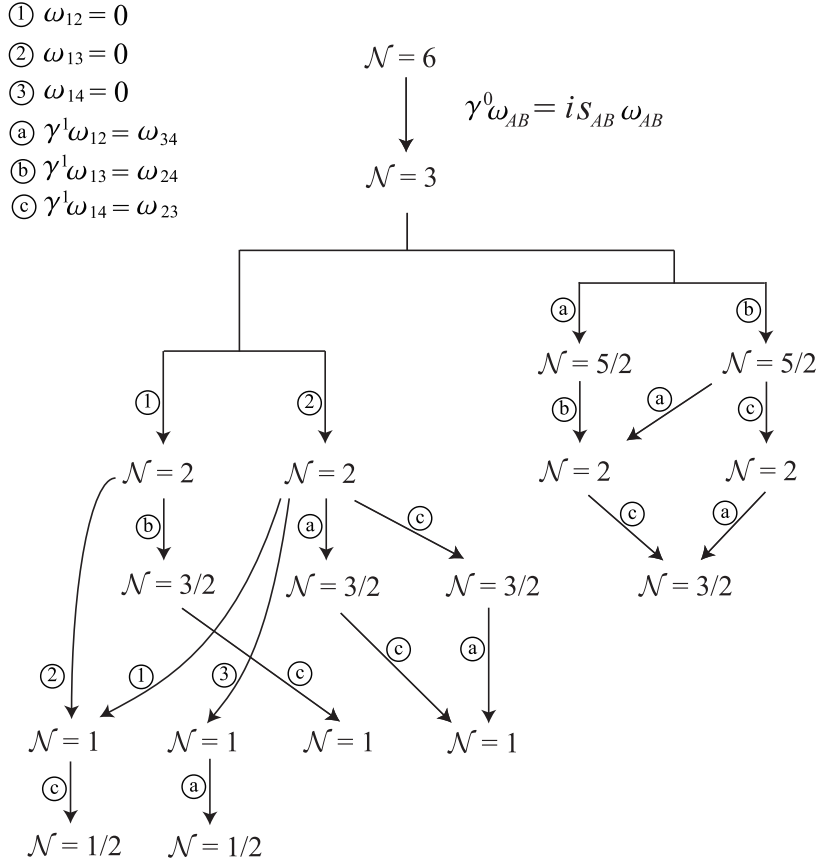}
} }
\caption{\small Supersymmetric cases for the vortex-type field configurations
}
\label{SUSYfig}
\end{figure}
\hspace{-6mm}

The action \eqref{ABJMa} possesses an SU(4) R-symmetry
and charge density for the SU(4) rotations are given by
\begin{equation}
J_{ab}^0 = i\left[Y^A (T_{ab})_A^{\;B} D_0 Y_B^\dag - D_0
Y^A(T_{ab})_A^{\;B} Y_B^\dag \right],
\end{equation}
where $T_{ab}$'s ($a,b=1,2,3,4$ and $a\neq b$) are six generators of SU(4) Lie algebra.
The mass deformation \eqref{delmpsi}
breaks the SU(4) R-symmetry to SU(2)$\times$SU(2)$\times$U(1) of which
the first SU(2) rotates ($Y^1$, $Y^2$), the second SU(2) does
($Y^3$, $Y^4$), and the U(1) transforms ($Y^1$, $Y^2$) and ($Y^3$,
$Y^4$) with opposite phases. For instance, consider an SU(2) rotation transforming $Y^1
\rightarrow e^{-i\alpha}Y^1$ and $Y^2 \rightarrow e^{i\alpha}Y^2$, and then
the corresponding R-charge is given by
\begin{align}
R_{12} = \int d^2x\,{\rm tr}\, J_{12}^0 =\int d^2x\,{\rm
tr}\left[i(Y^1 D_0Y_1^\dagger - D_0 Y^1 Y_1^\dagger)
         - i(Y^2 D_0 Y_2^\dagger - D_0 Y^2 Y_2^\dagger)\right].\label{r12}
\end{align}

In order to obtain the vortex-like BPS configurations, we will impose some supersymmetric conditions
to the supersymmetric parameters $\omega^{AB}$,  which reduces the number of supersymmetries.
Possible supersymmetries and the corresponding supersymmetric conditions
are depicted schematically in Fig.~\ref{SUSYfig}.

\setcounter{equation}{0}
\section{Vortex-type Objects with ${\cal N}=3$ Supersymmetry }\label{sec3}

The vortex-type half-BPS solitons have been discussed in
\cite{Kim:2009ny}. In this section we briefly summarize
the result of our previous work and discuss more general solutions in
mass-deformed case.

\subsection{BPS equations and bound}\label{ss31}

Supersymmetric variation of the fermion field $\psi_A$ in \eqref{strf} is,
\begin{align}\label{kspr}
0&=\gamma^0\delta \psi_A\nonumber\\&=\left[-\delta_A^{[B} D_0 Y^{C]}
          + \gamma^0\left( \beta^{BC}_{\;A} + \delta^{[B}_A \beta^{C]D}_{\;D}
          + \mu M_A^{\;[B} Y^{C]} \right)\right]\omega_{BC}-\gamma^2(D_1 -\gamma^0 D_2)Y^B\omega_{AB}.
\end{align}
Now we impose the supersymmetric condition
$\gamma^{0}\omega_{AB}=is_{AB} \omega_{AB}$ with $s_{AB} = s_{BA} =
\pm1$ to the equations (\ref{kspr}), which reduces the number of
supersymmetries by half. Then we have the following BPS equations
\cite{Kim:2009ny}:
\begin{align}\label{halfbps0}
(D_1 - is_{AB} D_2)Y^B &= 0, \nonumber \\
\delta_A^{[B} D_0 Y^{C]}
          -is_{BC} \left( \beta^{BC}_{\;A} + \delta^{[B}_A \beta^{C]D}_{\;D}
          + \mu M_A^{\;[B} Y^{C]} \right) &= 0,\qquad
\text{(no sum over $B,C$)}.
\end{align}
More explicitly, assuming that $Y^1$ is nontrivial, we have
\begin{alignat}{3}\label{hpeq}
&D_1 Y^1-isD_2 Y^1=0,\qquad &&D_1 Y^B=D_2 Y^B=0 \qquad (B\neq1),\nonumber\\
&D_0 Y^1 +is (\beta^{21}_{\;2} + \mu Y^1)  = 0, \qquad
&&D_0 Y^2 -is (\beta^{12}_{\;1} + \mu Y^2)  = 0, \nonumber \\
&D_0 Y^3 -is \beta^{13}_{\;1} = 0, \qquad
&&D_0 Y^4 -is \beta^{14}_{\;1} = 0, \nonumber \\
&\beta^{31}_{\;3} = \beta^{41}_{\;4} = \beta^{21}_{\;2} + \mu Y^1,
\qquad &&\beta^{43}_{\;4} = \mu Y^3, \qquad
\beta^{34}_{\;3} = \mu Y^4, \nonumber \\
&\beta^{32}_{\;3} = \beta^{42}_{\;4} =
\beta^{23}_{\;2} = \beta^{24}_{\;2} = 0,&& \nonumber \\
&\beta^{BC}_{\;A} = 0 \qquad(A \neq B \neq C \neq A),&&
\end{alignat}
where $s=\pm1$.

The BPS equations \eqref{halfbps0} can also be obtained from
energy expression \eqref{Ene},
\begin{align} \label{ene32}
E &= \frac13 \int d^2x \,{\rm tr}\, \left\{ 2\sum_{A,B,C} \left|
\delta_A^{[B} D_0 Y^{C]}
          -is_{BC} \left( \beta^{BC}_{\;A} + \delta^{[B}_A \beta^{C]D}_{\;D}
          + \mu M_A^{\;[B} Y^{C]} \right) \right|^2 \right.
\nonumber\\
& \hspace{25mm}\left.
     + \sum_{A\neq B}|(D_1 - is_{AB} D_2)Y^A|^2 \right\}
\nonumber \\
  &\hspace{5mm} + is\, {\rm tr} \int d^2x \epsilon_{ij} \partial_i \left(
       Y_1^\dagger D_j Y^1 - \frac13 \sum_{A=2}^4 Y_A^\dagger D_j Y^A \right)
   - \frac{s}3\, \mu\, {\rm tr}
                \int d^2x ( j^0 + 2 J_{12}^0 ).
\end{align}
For any well-behaved BPS configuration satisfying the BPS equations
(\ref{hpeq}), the energy is bounded by both the U(1) charge
(\ref{u1c}) and the R-charge (\ref{r12}),
\begin{equation} \label{ene322}
E \ge  \frac13\, |\mu (Q + 2R_{12})|.
\end{equation}

One can also reshuffle the stress components of the energy momentum
tensor \eqref{eneT},
\begin{align}
T_{ij}=&\frac{2}{3}\,\eta_{ij}\, {\rm Re\, tr}\Bigg\{
 \left[ \delta_A^{[B} D_0 Y^{C]}
        +is_{BC} ( \beta^{BC}_{\;A} + \delta^{[B}_A \beta^{C]D}_{\;D}
        + \mu M_A^{\;[B} Y^{C]})\right]^\dag \notag\\
&\hspace{24mm}\times\left[
 \delta_A^{[B} D_0 Y^{C]}
  -is_{BC} ( \beta^{BC}_{\;A} + \delta^{[B}_A \beta^{C]D}_{\;D}
  + \mu M_A^{\;[B} Y^{C]} ) \right]  \Bigg\} \notag\\
& + \frac12 {\rm Re\, tr} \left\{
 [(D_i+is\epsilon_{ik} D_k) Y^A ]^\dag (D_j-is\epsilon_{jl} D_l) Y^A
 + ( i \leftrightarrow j ) \right\},
\end{align}
which clearly vanish if the BPS equations \eqref{halfbps0} are imposed.
Note that from the spatial
component of the energy-momentum conservation, the force density
${\cal{F}}^i$ at a given spacetime point $(t,x^i)$ is
\begin{align}\label{forc}
{\cal F}^i = \frac{\partial}{\partial t} T^{i0}=\nabla_j T^{ij}.
\end{align}
Thus, vanishing $T_{ij}$ is a sufficient condition of vanishing
force everywhere and any static multi-BPS solitons (or
anti-solitons) with vanishing $T_{ij}$ are noninteracting, at least
at the classical level.

\subsection{BPS objects in the massless theory}\label{ss33}
For the original ABJM theory without mass deformation ($\mu = 0$),
it has been shown that the BPS equation \eqref{hpeq} is
equivalent to \cite{Kim:2009ny}
\begin{align} \label{reduced}
(D_1 - is D_2)Y^1 &= 0, \nonumber \\
Y^A &= v^A I, \qquad (A=2,3,4), \nonumber \\
B = \hat B &= -\frac{s}2 \left( \frac{2\pi v}{k} \right)^2
[Y^1,Y_1^\dagger],
\end{align}
where $v^A$ ($A=2,3,4$) are constants and $v^2 = \sum_{A=2}^4
|v^A|^2$. Note that all the constraints in \eqref{hpeq} are
completely solved. The Eq.~\eqref{reduced} is nothing but the Hitchin equation
\cite{Hitchin:1986vp} which is a half-BPS equation of super Yang-Mills
theory with the identification $g_\textrm{YM}=\frac{2\pi v}{k}$. This
identification has appeared in the context
of the compactification of ABJM theory (from M2 to D2)
\cite{Mukhi:2008ux,Pang:2008hw,Kim:2011qv}.
See also Refs.~\cite{Go:2011bs,Jeon:2012fn,Kim:2012gz}.

Under a suitable ansatz \cite{Kim:2009ny}, Eq.~\eqref{reduced} is
reduced to (affine-) Toda-type equation,
\begin{align}\label{udeqn}
\partial \bar\partial \ln |y_a|^2 &= 4v \left(\frac{2\pi}{k}\right)^2
\sum_{b=1}^{N-1} K_{a b}\left( |y_b|^2 -
\frac{|G(z)|^2}{|c_b|^2\prod_{c=1}^{N-1} |y_c|^2}\right),
\nonumber \\
y_M &= \frac{G(z)}{\prod_{a=1}^{N-1} y_a},
\end{align}
where $G(z)$ is an arbitrary holomorphic function. For SU(2), this
becomes to Liouville-type equation (with $G=0$) or Sinh-Gordon-type
equation (with $G=$const.).

\subsection{BPS objects in the mass-deformed theory}\label{ss34}

In the mass-deformed theory ($\mu \neq 0$), the constraint equations
in \eqref{hpeq} have not been solved completely in general except $N=2,3$.
Here we briefly summarize U(2)$\times$U(2) case discussed in \cite{Kim:2009ny}
and generalize the result to U($N)\times$U($N$) case.

Solving the constraints in \eqref{hpeq}, it turns out that
scalar fields can be written in the form
\begin{align} \label{u2ya}
Y^1 &= \sqrt{\frac{k\mu}{2\pi}}
       \begin{pmatrix} 0 & f \\ 0 & 0 \end{pmatrix}, \qquad
Y^2 = \sqrt{\frac{k\mu}{2\pi}}
      \begin{pmatrix} a & 0 \\ 0 & \sqrt{a^2+1} \end{pmatrix}, \nonumber \\
Y^3 &= Y^4 = 0,
\end{align}
while the magnetic fields take the diagonal form
\begin{equation} \label{diagb}
B = \hat B = -2s \mu^2
    \begin{pmatrix} a^2 (1+|f|^2) & 0 \\ 0 & (a^2+1)(1-|f|^2) \end{pmatrix},
\end{equation}
where $a$ is a nonnegative constant.
Combining these two using the equations in the first line of \eqref{hpeq}
results in
\begin{equation} \label{mh}
\partial \bar\partial \ln |f|^2
+i(\partial{\bar \partial}-{\bar \partial}\partial )\Omega
=  \mu^2 \left[ (2a^2+1)|f|^2 -1 \right],
\end{equation}
where $\Omega$ is the phase of the scalar field, $f=|f|e^{i\Omega}$.
This is the well-known vortex equation appearing in Maxwell-Higgs theory.
Note that the phase of the scalar field $\Omega$ in two spatial dimensions
can be decomposed into a smooth part $\Omega_{{\rm reg}}$ and a singular part,
\begin{equation}
\Omega_{{\rm sing}}
=-\frac{i}{2}\,{\rm ln}\,\prod_{p=1}^{n}\frac{z-z_p}{\bar{z}-\bar{z_p}}.
\label{Osi}
\end{equation}
This gives the 2-dimensional Green's function
\begin{equation}
i(\partial\bar\partial-\bar\partial\partial)\Omega
=i(\partial\bar\partial-\bar\partial\partial)\left(\Omega_{\rm
reg}+\Omega_{\rm sing}\right) =-\frac{1}{4}\nabla^2\,{\rm ln}
\prod_{p=1}^n |z-
z_p|^2=-\pi{\sum_{p=1}^n}\delta(\vec{x}-\vec{x}_p), \label{omerel}
\end{equation}
where $n$ is interpreted as vorticity of
multi-vortex configurations.

The energy of the solution is a sum of two terms,
\begin{equation}
E = \frac{nk\mu}{2a^2+1}
        +  \left|\frac{k\mu}{2\pi}B_0\, \mathrm{tr} \int d^2x\, \right|,
\end{equation}
where $B_0=-4s\mu^2 a^2 (a^2+1)$ is the asymptotic value of the magnetic
field in \eqref{diagb}. Therefore solutions with nonzero $a$ may be
interpreted as vortices in a constant magnetic field.

Considering the vacuum configurations \eqref{gvac},
we can generalize the ansatz \eqref{u2ya}
(with $a=0$) to that of U($N)\times{\rm U}(N)$ case,
\begin{align}\label{N=3UN}
Y^{1}&=\sqrt{ \frac{k \mu}{2\pi} }\left(\begin{array}{c}
\begin{array}{cccccc}{\bf 0}_{m_1\times (m_1+1)}&&&&&\\&\ddots&&&&\\
&&{\bf 0}_{m_I\times(m_I + 1)} &&&\\
&&& f_1{\mathcal M^\dag}_1^{(m_{I+1})}\!\!&&\\&&&&\!\!\ddots\!&\\
&&&&&\!\!{f_K\mathcal M^\dag}_1^{(m_{I+K})}\end{array}\\
\end{array}\right),
\end{align}
where $f_k$'s ($k=1,\cdots,K$) are arbitrary complex functions.
We fix the remaining complex scalar fields $Y^{2,3,4}$
as the vacuum configurations given in \eqref{gvac}.
This ansatz satisfies all constraints in \eqref{hpeq}.
Combining the Gauss constraints \eqref{Beq}, \eqref{Bheq}
and the first order differential equations in the second and the third lines of \eqref{hpeq},
one can reduce the gauged Cauchy-Riemann equation of $Y^1$ in \eqref{hpeq} to the second order
differential equations,
\begin{align}
&\partial\bar\partial \,{\rm
ln}\frac{|f_k|^2}{\displaystyle{\prod_{p=1}^{n_{k}}}|z-z_p^{(k)}|^{2}}
=\mu^2(|f_k|^2-1),\qquad (k=1,\cdots,K),
\end{align}
where $n_k$ and $z_p^{(k)}$ denote the vorticity and the position of
the zeroes of $f_k$, respectively.
Its energy is given by
\begin{align}
&E=k\mu{\displaystyle\sum_{k=1}^K}\frac{m_k(m_k-1)n_{k}}{2}.
\end{align}

It is worth noting that the angular momentum \eqref{anm} of the solution
vanishes contrary to the usual spinning BPS vortices in
Chern-Simons Higgs theory ~\cite{Hong:1990yh, Jackiw:1990pr}. This is because
fields do not carry both charge and vorticity, i.e., either $D_0Y^A$ or
$D_iY^A$ vanishes in this case. In the next section, however, we will
see that solutions with less supersymmetries have nonzero angular momenta.

\setcounter{equation}{0}
\section{Vortex-type Objects with $\cal N$=2 Supersymmetry }\label{sec4}

In this section we consider vortex-type ${\cal N}=2$ BPS solitons without
and with mass deformation.
As we see in Fig.1, we can obtain ${\cal N}=2$ configurations by imposing
one of the following conditions

(i) $\omega_{12} =  0$,

(ii) $\omega_{13} =0$,

(iii) $\gamma^1\omega_{12} = \omega_{34},\,\,
\gamma^1\omega_{14}= \omega_{23}$,

(iv) $\gamma^1\omega_{13} = \omega_{24},\,\,
\gamma^1\omega_{14}= \omega_{23}$

\noindent
in addition to the condition $\gamma^{0}\omega_{AB} = is_{AB}\omega_{AB}$
($s_{AB}=\pm 1$). Here we only consider the case (i).
We will treat the case (ii) in Appendix A.1.
In the undeformed ABJM theory, the cases (i) and (ii) are equivalent
due to the SU(4) R-symmetry. However, in the mass-deformed case, they are
inequivalent in general.
As discussed in section \ref{halfint}, the BPS solutions of cases
(iii) and (iv) are equivalent to those of ${\cal N}=3$ BPS equations.

The brane interpretations of the cases (i) and (ii),
which are equivalent in the massless case, were given
in section 5.3 of the Ref.~\cite{Fujimori:2010ec}.
These cases are interpreted as the configuration of intersecting M2-branes
spanning two complex coordinates. If we assume that the intersecting
M2-branes span only one complex, then the corresponding configuration
becomes that of ${\cal N}=3$ BPS equations discussed in the previous section.

\subsection{BPS equations and bound}\label{ss41}

When $\omega_{12}=0$,  the Killing spinor equation (\ref{kspr})
leads to the following BPS equations:
\begin{alignat}{3}
& (D_{1}- is D_{2})Y^{1}=0, \quad && (D_{1}+is D_{2})Y^{2}=0,
\nonumber\\
& D_{1}Y^{p}=D_{2}Y^{p}=0,~~(p=3,4), &&
\nonumber\\
& D_{0}Y^{1}+is(\beta^{21}_{\; 2}+\mu Y^{1})=0, \qquad &&
D_{0}Y^{2}-is(\beta^{12}_{\; 1}+\mu Y^{2})=0,
\nonumber\\
& D_{0}Y^{p}+is(\beta^{2p}_{\; 2}- \beta^{1p}_{\; 1}) =0,
&&
\nonumber\\
& \beta^{3a}_{\; 3}=\beta^{4a}_{\; 4}~~(a=1,2), \quad &&
\beta^{43}_{\; 4}-\mu Y^{3}=\beta^{34}_{\; 3}-\mu Y^{4}=0,
\nonumber\\
& \beta^{23}_{\; 1}=\beta^{24}_{\; 1}=\beta^{13}_{\;
2}=\beta^{14}_{\; 2} =\beta^{14}_{\; 3} = \beta^{24}_{\; 3}=
\beta^{13}_{\; 4} &&=\beta^{23}_{\; 4}=0.
\label{N22}
\end{alignat}
Compared with the $\mathcal{N}=3$ BPS equations \eqref{hpeq},
the main difference
is that we have nontrivial equations for $Y^2$:
a gauged Cauchy-Riemann equation
of $Y^2$ field and some constraint equations involving $Y^2$.
We expect that the $Y^2$ field is allowed to
have some nontrivial configurations instead of vacuum configurations
in the $\mathcal{N}=3$ BPS case \eqref{hpeq}. The obtained BPS objects
would be in general different from the Maxwell-Higgs type vortices of
the half BPS case. In special case (constant $Y^2$), they would reduce
to the ${\cal N}=3$ BPS objects discussed in the previous section.

As we did in section 3, we can obtain the energy bound by reshuffling terms
in the energy expression,

\begin{align}
E =& {\rm tr} \int d^2x  \,\left[
\sum_{a,p,q} \left| \delta_p^{[q} D_0 Y^{a]} -is_{qa}
\left( \beta^{qa}_{\;p} + \delta^{[q}_p \beta^{a]A}_{\;A} + \mu
M_p^{\;[q} Y^{a]} \right) \right|^2
+(1,2 \leftrightarrow 3,4) \right. \notag \\
&\hspace{16mm}+
\left|(D_1-isD_2)Y^1\right|^2+\left|(D_1+isD_2)Y^2\right|^2\notag\\
&\hspace{16mm}+\frac{1}{2}\sum_{p=3,4}\left(\left|(D_1-isD_2)Y^p\right|^2
         +\left|(D_1+isD_2)Y^p\right|^2\right)\Bigg] \notag\\
& +is \,{\rm tr}\int
d^2x\epsilon_{ij}\partial_{i}(Y_1^\dag D_j Y^1-Y_2^\dag D_j
Y^2)-s\mu\,{\rm tr} \int d^2x \,J^0_{12}, \label{ene51}
\end{align}
where $a=1,2$ and $p,q=3,4$.
By requiring the square terms to vanish, we reproduce
the BPS equations \eqref{N22}.
The first term in the last line is a boundary term which vanishes for any
well-behaved field configuration. Note that in the mass-deformed theory with
$\mu\ne 0$, unlike the half BPS case in \eqref{ene322},
the energy is not bounded by the ${\rm U}(1)$ charge $Q$ but
the global SU(2) R-charge $R_{12}$
\eqref{r12},
\begin{equation} \label{ene2}
E \ge  | \mu {\rm tr}R_{12}|.
\end{equation}

The stress components of energy-momentum tensor \eqref{eneT} are
also written as
\begin{align}
T_{ij}&=\eta_{ij}\, {\rm Re\, tr} \Bigg\{
\left[
\delta_p^{[q}D_0Y^{a]} +is_{qa} ( \beta^{qa}_{\;p} + \delta^{[q}_p
\beta^{a]A}_{\;A} + \mu M_p^{\;[q} Y^{a]}
)\right]^\dag\nonumber\\
&\hspace{22mm}\times\left[ \delta_p^{[q} D_0
Y^{a]} -is_{qa} ( \beta^{qa}_{\;p} + \delta^{[q}_p \beta^{a]A}_{\;A}
+ \mu M_p^{\;[q} Y^{a]})\right]
+ (1,2 \leftrightarrow 3,4) \Bigg\} \notag \\
&+\frac12 {\rm Re\, tr}\,\Big\{ [(D_i +is\epsilon_{ik}
D_k) Y^A]^\dag(D_j -is\epsilon_{jl} D_l)Y^A
+(i \leftrightarrow j)
\Big\},
\label{Str2}
\end{align}
which vanishes everywhere on imposing
${\cal N}=2$ BPS equations \eqref{N22}.
As discussed in $\eqref{forc}$ this pointwise
absence of force guarantees that the obtained BPS objects are
classically noninteracting.

\subsection{BPS objects in the massless theory}\label{ss42}

In the massless case ($\mu=0$), the energy \eqref{ene51} is
bounded by the total derivative term,
as already discussed in the massless half-BPS case,
\begin{align}
E=\left|is \,{\rm tr}\int d^2x\epsilon_{ij}\partial_{i}(Y_1^\dag D_j
Y^1-Y_2^\dag D_j Y^2)\right|,
\end{align}
which vanishes for any well-behaved field configuration. In this case, we expect
that there is no regular soliton solution with finite energy.

With a U($N$)$\times$U($N$) gauge transformation, we may assume
without loss of generality that $Y^3$ is diagonal,
\begin{equation} \label{y3}
Y^3 = \begin{pmatrix}
            v^3_1 I_{n_1} &             &        &             \\
                        & v^3_2 I_{n_2} &        &             \\
                        &             & \ddots &             \\
                        &             &        & v^3_k I_{n_k}
      \end{pmatrix}, \qquad
(0 \le v^3_1 < v^3_2 < \cdots < v^3_k).
\end{equation}
From the constraints $\beta^{34}_{\;3}=\beta^{43}_{\;4}=0$
(with $\mu=0$) in \eqref{N22}, $Y^4$ has to be also diagonal.
Applying the other constraints $\beta^{14}_{\;3}=\beta^{24}_{\;3}=0$ in
\eqref{N22}, we notice
that $Y^1$ and $Y^2$ are block diagonal and, in each block diagonal subspace,
$Y^4$ should be proportional to the identity.
Then, for each subspace where $Y^A=v^A I\,\, (A=3,4)$,
nontrivial BPS equations in \eqref{N22} become
\begin{align}\label{2BP}
&(D_1 - is D_2)Y^1 = 0, \qquad (D_1 + is D_2)Y^2 = 0, \nonumber \\
&B = -2s \left( \frac{2\pi}{k} \right)^2 \left\{[Y^1 Y_2^\dag,Y^2
Y_1^\dagger]+v^2\left([Y^1,Y_1^\dag]-[Y^2,Y_2^\dag]\right)\right\},\nonumber\\
&\hat B = -2s \left( \frac{2\pi}{k} \right)^2 \left\{[Y_2^\dag
Y^1,Y_1^\dagger
Y^2]+v^2\left([Y^1,Y_1^\dag]-[Y^2,Y_2^\dag]\right)\right\},
\end{align}
where $v^2=\sum_{A=3,4} |v^A|^2$. When one of $Y^{1}$ and $Y^{2}$ is
assumed to be proportional to identity in each block diagonal
subspace, \eqref{2BP} reduce to \eqref{reduced} in the $\mathcal{N}=3$ BPS case.

We consider some simple solutions of $\eqref{2BP}$ with $s=1$ for
definiteness. For U(2)$\times$U(2) case, we take an ansatz,
\begin{equation}\label{ann2}
Y^1 =
      \begin{pmatrix} 0 & d \\ e & 0 \end{pmatrix},\qquad
Y^2 =
      \begin{pmatrix} a & 0 \\ 0 & b \end{pmatrix}.
\end{equation}
Plugging \eqref{ann2} into $\eqref{2BP}$ we obtain
\begin{align}
&ab=C,\qquad de=D,\nonumber\\
&\partial\bar{\partial}\,{\rm{ln}} \,
|b|^2-i(\partial\bar{\partial}-\bar{\partial}\partial)\Omega_b
=\left(\frac{2\pi}{k}\right)^2(|b|^2-|a|^2) (|d|^2+|e|^2),
\label{eqn2-1}\\
&\partial\bar{\partial}\, {\rm{ln}}
\,|d|^2-i(\partial\bar{\partial}-\bar{\partial}\partial)\Omega_d
=\left(\frac{2\pi}{k}\right)^2(|a|^2+|b|^2+2v^2)(|d|^2-|e|^2),
\label{eqn2-2}
\end{align}
where $\Omega$'s are phases of scalar fields, $b=|b|e^{-i\Omega_b}$,
$d=|d|e^{i\Omega_d}$ and $C, D$ are arbitrary constants.
For $Y^2=I$, $a=b$, \eqref{eqn2-1} and
\eqref{eqn2-2} are reduced to the Liouville-type equation (with $D = 0$) or
Sinh-Gordon-type equation (with $D =$ const) which we obtained in
subsection~\ref{ss33} as $\mathcal{N}=3$ BPS configurations.
On the other hand, with $a=e=v=0$, the equations are further simplified to
\begin{align} \label{n2l}
\partial\bar{\partial}\,{\rm{ln}} \,
|b|^2-i(\partial\bar{\partial}-\bar{\partial}\partial)\Omega_b
&=\left(\frac{2\pi}{k}\right)^2|b|^2 |d|^2, \notag \\
\partial\bar{\partial}\, {\rm{ln}}
\,|d|^2-i(\partial\bar{\partial}-\bar{\partial}\partial)\Omega_d
&=\left(\frac{2\pi}{k}\right)^2|b|^2 |d|^2,
\end{align}
which again become a Liouville equation with $b=d$.

\subsection{BPS objects in the mass-deformed theory}

\subsubsection{U(2)$\times$U(2) gauge group}

Let us first consider the simplest U(2)$\times$U(2) case.
By the same reasoning developed in \cite{Kim:2009ny} to obtain
reduced equations in $\mathcal{N}=3$ BPS case \eqref{mh}, it is readily shown
that the constraint equations in \eqref{N22} lead us to put $Y^3=Y^4=0$
for nontrivial solutions. Then we are left with
\begin{alignat}{2}
(D_1+iD_2) Y^1&=0, & (D_1-iD_2) Y^2&=0, \notag\\
D_0 Y^1-i(\beta^{21}_{\;2}+\mu Y^1)&=0, \qquad &
D_0 Y^2+i(\beta^{12}_{\;1}+\mu Y^2)&=0,
\label{B21}
\end{alignat}
as well as the Gauss' laws \eqref{Beq} and \eqref{Bheq}.
Comparing with the half-BPS case, we have nontrivial equations
for $Y^2$ in addition to $Y^1$ and they are coupled to each other.

We proceed by adopting a simple ansatz
from the broken vacuum \eqref{gvac}. More specifically,
in U(2)$\times$U(2) case, we consider
\begin{equation}\label{Yan2}
Y^1 = \sqrt{ \frac{k \mu}{2\pi} }
      \begin{pmatrix} 0 & d \\ 0 & 0 \end{pmatrix},\qquad
Y^2 = \sqrt{ \frac{k \mu}{2\pi} }
      \begin{pmatrix} 0 & 0 \\ 0 & b \end{pmatrix},\qquad Y^3=Y^4=0,
\end{equation}
which is actually the same ansatz used in the previous section to obtain
\eqref{n2l}. Moreover, comparing with the ansatz \eqref{u2ya} employed
in half-BPS case, we see that \eqref{Yan2} has essentially the same
form as \eqref{u2ya} (with $a=0$). Here, thanks to the less supersymmetries,
$Y^2$ is no longer a constant. We will see below that this freedom
allows us to have richer solutions with nonvanishing angular momenta.

With the above ansatz, the Gauss' laws \eqref{Beq} and \eqref{Bheq}
take simple diagonal forms,
\begin{align}\label{B1}
B=2 \mu^2\begin{pmatrix}
  -|d|^2(1-|b|^2) & 0\\ 0 &
  |b|^2(1-|d|^2)
  \end{pmatrix},\qquad
{\hat B}=2\mu^2\begin{pmatrix}
  0 & 0\\ 0 &
  |b|^2-|d|^2\end{pmatrix},
\end{align}
from which we can write the corresponding gauge fields as
\begin{align}\label{AAH}
A= \begin{pmatrix}
  u & 0\\ 0 &
  v
  \end{pmatrix},\qquad
{\hat A}=\begin{pmatrix}
  0 & 0\\ 0 &
  \hat v \end{pmatrix},
\end{align}
where
\begin{align}\label{Ba}
B=\frac{2}{i}(\partial {\bar A} - \bar{\partial}
A),\qquad
\hat B=\frac{2}{i}({\partial} \bar{\hat A} - \bar{\partial} {\hat
A}).
\end{align}
Substituting \eqref{AAH} and \eqref{Yan2} into the first line of BPS equations \eqref{B21} we have
\begin{equation}\label{uvbd}
{\bar u} -{\bar {\hat v}} = i {\bar \partial {\rm ln} d},\qquad v-{\hat v}=i \partial {\rm ln} b.
\end{equation}
Inserting \eqref{uvbd} into the magnetic field \eqref{Ba} and
comparing it with \eqref{B1}, we obtain two equations for scalar fields
\begin{align}
&\partial\bar\partial \,{\rm
ln}\frac{|b|^2}{\displaystyle{\prod_{p=1}^{n_b}}|z-z_p|^{2}}
=\mu^2|d|^2(|b|^2-1),\label{eom2-5}\\
&\partial\bar{\partial}\, {\rm{ln}} \frac{ |d|^2}{
{\displaystyle{\prod_{q=1}^{n_d}}|z-z'_q|^{2}}}=\mu
^2|b|^2(|d|^2-1). \label{eom2-6}
\end{align}
The energy bounded by the R-charge \eqref{ene2} is rewritten as
\begin{align}
&E=\frac{k\mu^3}{\pi}\int d^2 x
\left[|b|^2(1-|d|^{2})+|d|^2(1-|b|^2)\right].
\label{ene53}
\end{align}

The reduced equations \eqref{eom2-5} and \eqref{eom2-6} have been considered in
self-dual $U(1)\times U(1)$ Chern-Simons system with two scalar fields
\cite{Kim:1993mh}, which may be considered as the abelian part of the
theory under consideration. Let us analyze the equations in the present setting.
From the expression of the energy \eqref{ene53}, it follows that there are
two classes of boundary conditions at spatial infinity,
\begin{align}
|b(\infty)|=|d(\infty)|=1,\qquad |b(\infty)|=|d(\infty)|=0,
\label{bcin}
\end{align}
which correspond to topological and nontopological solutions, respectively.
At every vortex point $z=z_p,\ z'_q$, singlevaluedness of the fields
requires their amplitudes to vanish, $|b(z_p)|=|d(z'_q)|=0$.
(If $n_b=0$ or $n_d=0$, the corresponding amplitudes need not vanish.)
Depending on the vorticity, nontopological solutions can further be
classified as nontoplogical $Q$-balls ($n_{b}=n_{d}=0$), $Q$-vortices
($n_{b}\ne 0$ and $n_{d}\ne0$), and their hybrids ($n_{b}\ne 0$ and $n_{d}=0$).

These gauged vortices carry diagonal components of magnetic fluxes of which
the contributions come from the spatial infinity,
\begin{align}
\Phi=&4\oint_{|x^{i}|\rightarrow\infty}\hspace{-3mm} dx^{i}
\begin{pmatrix}\partial_{i}
\displaystyle{\ln\frac{|b|}{\prod_{p=1}^{n_b}|x-x_{p}|}} & 0\\
0 & \displaystyle{-\partial_{i}
\ln\frac{|d|}{\prod_{q=1}^{n_d}|x-x_{q}|}}
 \end{pmatrix},
\label{pBP}\\
\hat \Phi=&4\oint_{|x^{i}|\rightarrow\infty}\hspace{-3mm} dx^{i}
 \begin{pmatrix}
0 & 0\\
0 &  \displaystyle{\partial_{i}
\ln\frac{|b|\prod_{q=1}^{n_d}|x-x_{q}|}{|d|
\prod_{p=1}^{n_b}|x-x_{p}|}}
 \end{pmatrix}.
\label{phBP}
\end{align}

We parameterize the asymptotic behaviors of the fields as
\begin{align}
|b|\sim |x^{i}|^{-\alpha_{b}}, \qquad
|d|\sim |x^{i}|^{-\alpha_{d}},
\label{bdaa}
\end{align}
where $\alpha_b$ and $\alpha_d$ are positive constants for nontopological
solutions, while $\alpha_b = \alpha_d =0$ for topological ones.
Then the magnetic fluxes \eqref{pBP}--\eqref{phBP} become
\begin{align}
\Phi=&2 \pi\begin{pmatrix}
   -n_b-\alpha_b & 0\\ 0 &
   n_d+\alpha_d
  \end{pmatrix},\qquad  {\hat \Phi}=2 \pi\begin{pmatrix}
   0 & 0\\
   0 & -n_b-\alpha_b+ n_d+\alpha_d
  \end{pmatrix}.
\label{flx}
\end{align}
For the toplogical vortices satisfying the first boundary condition in
\eqref{bcin}, the fluxes are quantized by the integer-valued vorticities
$n_{b}$ and $n_{d}$, as $\alpha_{b}$ and $\alpha_{d}$ are zero.

In Chern-Simons gauge theories,
the Gauss' laws \eqref{Beq}--\eqref{Bheq} with the help of
the conserved currents \eqref{Ucu} imply that
\begin{align}
Q=\frac{k}{2\pi} \Phi,\qquad \hat Q=\frac{k}{2\pi} \hat \Phi,
\label{QQ}
\end{align}
and hence the flux carrying objects are also charged.
Since the energy of $\mathcal{N}=2$ BPS solitons is bounded by the trace of
R-charge \eqref{ene51}--\eqref{ene2}, they carry R-charge \eqref{r12} as well,
\begin{align}
R_{12}=& \frac{k\mu^2}{\pi} \int d^{2}x \begin{pmatrix}
  |d|^2(1-|b|^2) & 0\\ 0 &
  |b|^2(1-|d|^2)
  \end{pmatrix}\nonumber\\
=&k\begin{pmatrix}
   n_b+\alpha_b & 0\\ 0 &
   n_d+\alpha_d
  \end{pmatrix}.
\label{Rna}
\end{align}

As we mentioned above, a notable difference from the $\mathcal{N}=3$ BPS case is
that the solution carries nonzero angular momentum in the present case.
In this regard, note in particular that both $D_0 Y^a$ and $D_i Y^a$ ($a=1,2$)
are not zero from the BPS equations \eqref{B21}.
Therefore the angular momentum \eqref{anm} does not vanish,
\begin{align}
J=-\frac{k}{2\pi}\int d^{2}x\, \epsilon_{ij} x_i \Big[  (v_j
-\partial_j \Omega_d)B_{11}
+ (u_j +\partial_j \Omega_b)B_{22}\Big].
\label{Jk}
\end{align}
The explicit value of $J$ can be computed for rotationally symmetric
solutions as seen below.

For rotationally symmetric configurations we
take the ansatz
\begin{equation}
\Omega_b=n_b \theta,\qquad \Omega_d=n_d \theta,
\label{Obd}
\end{equation}
as well as
\begin{align}
u_i=-\epsilon_{ij}\frac{x^j}{r^2} u_r(r),\qquad
v_i=-\epsilon_{ij}\frac{x^j}{r^2} v_r(r).\label{uvir}
\end{align}
Then after a straightforward calculation \label{appendix_c} (see also
\cite{Kim:1993mh}) we find
\begin{align}
J= k (\alpha_b\alpha_d-n_bn_d).
\label{JJ}
\end{align}

For topological vortices,
the U(2)$\times$U(2) gauge symmetry is spontaneously broken to
U(1)$\times$U(1).
Since the fundamental group of the vacuum manifold is computed as
$\pi_{1}({\rm U}(2)\times {\rm U}(2)/{\rm U}(1)\times {\rm U}(1))=
\pi_{1}({\rm U}(1)\times {\rm S}^{2}\times {\rm U}(1)\times {\rm S}^{2})
={\mathbb Z}\times{\mathbb Z}$, the stability of the composite of
two static vortices is topologically guaranteed.
$Q$-balls and $Q$-vortices are generated in the symmetric phase of which
the vacuum has trivial topology and their stability should be examined
energetically~\cite{Friedberg:1976me}.
The mass of the transverse scalar fields $Y^{A}$ and that of the fermions
$\psi_{A}$ are all $\mu$ from \eqref{potential3} and \eqref{md2}.
Since the minimum energy to produce $Q$-balls and $Q$-vortices of R-charge
$R_{12}$ is given by \eqref{ene2}, the rest energy to produce the scalar or
fermion particles of the R-charge $R_{12}$ is exactly the same as the minimum
energy of $Q$-balls or $Q$-vortices of the same amount of R-charge. Therefore,
these $Q$-balls or $Q$-vortices are marginally stable~\cite{Jackiw:1990pr}
and this marginal
stability is a character of Chern-Simons Higgs theory in the BPS limit
with single mass scale.

In addition to rotationally symmetric solutions, we can obtain other class
of solutions of \eqref{eom2-5} and \eqref{eom2-6} for a few simple cases
\cite{Kim:1993mh}.
When the $|b|$ field takes the Higgs vacuum value $|b|=1$,
\eqref{eom2-5} becomes trivial and \eqref{eom2-6} reduces to the scalar
BPS equation for the Nielsen-Olesen type vortices
\begin{align}\label{MHv}
\partial\bar{\partial}\, {\rm{ln}} \frac{ |d|^2}{
{\displaystyle{\prod_{q=1}^{n_d}}|z-z_q|^{2}}}=\mu ^2(|d|^2-1),
\end{align}
which has already been discussed in the ${\cal N}=3$ case; see \eqref{mh}.
When $d$ and $b$ are parallel, \eqref{eom2-5}-\eqref{eom2-6} become single
scalar BPS equation~\cite{Arai:2008kv} equivalent to that for the
vortices in Abelian Chern-Simons-Higgs theory~\cite{Hong:1990yh}
\begin{equation}
\partial\bar{\partial}\, {\rm{ln}} \frac{ |d|^2}{
{\displaystyle{\prod_{q=1}^{n_d}}|z-z_q|^{2}}}=\mu
^2|d|^2(|d|^2-1).
\label{CII}
\end{equation}
This equation also supports the BPS multi-vortex-type solutions including
topological vortices~\cite{Hong:1990yh,Wang:1991na}
in the broken phase of $\displaystyle{\lim_{r
\rightarrow\infty}} |d|^2\rightarrow 1$, and $Q$-balls and
$Q$-vortices \cite{Spruck:1992yy} in the symmetric phase of
$\displaystyle{\lim_{r\rightarrow\infty}} |d|^2\rightarrow
0$~\cite{Jackiw:1990pr,Spruck:1992yy}.
In this case the magnetic fluxes \eqref{flx} are traceless so that
solutions do not carry U(1) magnetic fluxes, $\Phi_{{\rm U(1)}}=
{\hat \Phi}_{{\rm U(1)}}=0$ in \eqref{mfu1}, and U(1) charge,
$Q_{{\rm U(1)}}={\hat Q}_{{\rm U(1)}}=0$ in \eqref{u1c}.
It carries, however, a fractional angular momentum.
For rotationally symmetric configurations, it is given by
$J=k(\alpha_{d}^{2}-n_{d}^{2})$ from \eqref{JJ}.
When the topological vortices are separated from each other,
the scalar amplitude is expanded near a vortex point $z_{q}$ as
\begin{align}
\ln |d|^{2}\approx \ln |z-z_{q}|^{2}+a_{q}(z-z_{q})+ {\cal O}((z-z_{q})^{2}),
\end{align}
and then total angular momentum is given by a sum of spin part and orbital
part as
\begin{align}
J\approx -k|n_{d}|-k\sum_{q=1}^{n_{q}}|z_{q}a_{q}|.
\end{align}
For the sufficiently large separation, $a_{q}\rightarrow 0$ and there
remains only spin part linearly proportional to the
vorticity~\cite{Kim:1992yz}.

In addition, by evaluating the index of the differential
operator associated with the appropriate fluctuation equation,
one can find that the number of free parameters of the general solutions
of \eqref{eom2-5} and \eqref{eom2-6} \cite{Kim:1993mh} is given by
$2(n_b + n_d + [\alpha_b] + [\alpha_d])$, where $[\alpha_b]$ and
$[\alpha_d]$ respectively denote the largest integer less than $\alpha_b$
and $\alpha_d$.

\subsubsection{U(3)$\times$U(3) gauge group}

For higher-rank gauge groups, we proceed by setting $Y^3$ and $Y^4$ to
vacuum values. This is a natural choice considering that $D_i Y^p=0$ for all
$i=1,2$ and $p=3,4$ in the BPS equation \eqref{N22}, which
suggest no nontrivial dynamics
for $Y^{3,4}$.
Then from \eqref{gvac}, we have the following two possible
field configurations,
{\small\begin{align}
({\rm i})&\quad Y^3 = \sqrt{
\frac{k \mu}{2\pi} }
      \begin{pmatrix} 1 & 0 & 0\\ 0 & 0 & 0\\0 & 0 & 0\end{pmatrix},\quad
Y^4 = \sqrt{ \frac{k \mu}{2\pi} }
      \begin{pmatrix} 0 & 1 & 0\\ 0 & 0 & 0\\0 & 0 &
      0\end{pmatrix},\nonumber\\&\quad Y^1 = \sqrt{ \frac{k \mu}{2\pi} }
      \begin{pmatrix} 0 & 0 & 0\\ 0 & 0 & b\\0 & 0 & 0\end{pmatrix},\quad
Y^2 = \sqrt{ \frac{k \mu}{2\pi} }
      \begin{pmatrix} 0 & 0 & 0\\ 0 & 0 & 0\\0 & 0 &d\end{pmatrix},\nonumber\\
({\rm ii})&\quad Y^3 =Y^4=0,\quad Y^2 = \sqrt{ \frac{k \mu}{2\pi} }
      \begin{pmatrix} 0 & 0 & 0\\ 0 & b & 0\\0 & 0 &
      {\sqrt 2}b\end{pmatrix},\quad
      Y^1=\sqrt{ \frac{k \mu}{2\pi} }
      \begin{pmatrix} 0 & {\sqrt 2} d & 0\\ 0 & 0 & d\\0 & 0 &
      0\end{pmatrix}.\label{U3a}
\end{align}}

For the case (i), the magnetic field profiles are calculated as
\begin{align}\label{B33}
B= 2\mu^2 \begin{pmatrix} 0 &0&0\\0
&-|b|^2(1-|d|^2) &0\\0&0 &|d|^2(1-|b|^2)\end{pmatrix},\qquad \hat B= 2\mu^2
\begin{pmatrix} 0 &0&0\\0 &0&0\\0&0&|d|^2-|b|^2\end{pmatrix},
\end{align}
and then the resulting second order scalar BPS equations are
the same as those in ${\cal N}=2$ U(2)$\times$U(2) case,
\eqref{eom2-5}--\eqref{eom2-6}. The energy for these solitons, read
from the R-charge, is also given by \eqref{ene53}.

For the configuration (ii), we again end up with
\eqref{eom2-5}--\eqref{eom2-6}.
The energy and the angular momentum, however, have different values,
\begin{align}
E&= \frac{k\mu^3}{ \pi } {\rm tr} \int d^{2}x \begin{pmatrix}
  2|d|^2(1-|b|^2) & 0&0\\ 0 &
  |b|^2+|d|^2-2|b|^2|d|^2 &0 \\ 0 & 0 & 2|b|^2(1-|d|^2)
  \end{pmatrix} \nonumber\\
  &= 2k\mu\, {\rm tr} \begin{pmatrix}
   n_b+\alpha_b & 0 &0 \\ 0 & \frac12 (n_b+\alpha_b + n_d+\alpha_d) & 0
   \\ 0 & 0 & n_d+\alpha_d
  \end{pmatrix} \nonumber \\
 &= 3k\mu (n_b+\alpha_b + n_d+\alpha_d), \nonumber \\
J=& 3k(\alpha_b\alpha_d-n_b n_d),
\end{align}
where the angular momentum is calculated for rotational symmetric
configurations.

\subsubsection{U($N$)$\times$U($N$) gauge group}

For the U($N$)$\times$U($N$) gauge group, based on the vacuum configurations \eqref{gvac},
we consider the following field ansatz
\begin{align}\label{N=2UN}
Y^{1}&=\sqrt{ \frac{k \mu}{2\pi} }\left(\begin{array}{c}
\begin{array}{cccccc}{\bf 0}_{m_1\times (m_1+1)}&&&&&\\&\ddots&&&&\\
&&{\bf 0}_{m_I\times(m_I + 1)} &&&\\
&&& a_1{\mathcal M^\dag}_1^{(m_{I+1})}\!\!&&\\&&&&\!\!\ddots\!&\\
&&&&&\!\!{a_K\mathcal M^\dag}_1^{(m_{I+K})}\end{array}\\
\end{array}\right),\nonumber\\
Y^{2}&=\sqrt{ \frac{k \mu}{2\pi} }\left(\begin{array}{c}
\begin{array}{cccccc}{\bf 0}_{m_1\times (m_1+1)}&&&&&\\&\ddots&&&&\\
&&{\bf 0}_{m_I\times(m_I + 1)} &&&\\
&&& b_1{\mathcal M^\dag}_2^{(m_{I+1})}\!\!&&\\&&&&\!\!\ddots\!&\\
&&&&&\!\!{b_K\mathcal M^\dag}_2^{(m_{I+K})}\end{array}\\
\end{array}\right),
\end{align}
where $a_k$, $b_k$ $(k=1,...,K)$ are arbitrary complex functions,
and $Y^3$ and $Y^4$ are set to vacuum values. The resulting scalar
BPS equations are $K$ pairs of coupled differential equations
\begin{align}
&\partial\bar\partial \,{\rm
ln}\frac{|a_k|^2}{\displaystyle{\prod_{p=1}^{n_{a,k}}}|z-z_p|^{2}}
=\mu^2|b_k|^2(|a_k|^2-1),\qquad \partial\bar{\partial}\, {\rm{ln}}
\frac{ |b_k|^2}{ {\displaystyle{\prod_{q=1}^{n_{b,k}}}|z-z'_q|^{2}}}=\mu
^2|a_k|^2(|b_k|^2-1). \label{eom28}
\end{align}
Summing over all contributions from each block in \eqref{N=2UN},
we obtain the energy and angular momentum
\begin{align}
&E=k\mu{\displaystyle\sum_{k=1}^K}\frac{m_k(m_k-1)}{2} (n_{a,\,k}+\alpha_{a,\,k} + n_{b,\,k}+\alpha_{b,\,k}),\nonumber
\\ &J=k{\displaystyle\sum_{k=1}^K}\frac{m_k(m_k-1)}{2} (\alpha_{a,\,k}\alpha_{b,\,k}-n_{a,\,k} n_{b,\,k}),
\end{align}
where the angular momentum is calculated for the rotational symmetric
configuration as before.

When we choose $Y^2$  to the vacuum configuration by fixing $b_k=1$,
the second order differential equations \eqref{eom28} and the expressions of energy and angular momentum
are reduced to those of ${\cal N}=3$ case.
In this case the supersymmetry of the corresponding BPS configuration is
enhanced to ${\cal N}=3$.

\setcounter{equation}{0}
\section{Vortex-type Objects with $\cal N$=$1$ Supersymmetry}\label{sec5}

As we see in Fig.1, there are four ways to obtain ${\cal N}=1$ BPS equations:

(i) $\omega_{13} = \omega_{14} = 0$,

(ii) $\omega_{12} = \omega_{13}=0$,

(iii) $\omega_{13}=0,\,\, \gamma^1\omega_{12} = \omega_{34},\,\,
\gamma^1\omega_{14}= \omega_{23}$,

(iv) $\omega_{12}=0,\,\, \gamma^1\omega_{13} = \omega_{24},\,\,
\gamma^1\omega_{14}= \omega_{23}$

\noindent
in addition to the condition
$\gamma^{0}\omega_{AB} = is_{AB}\omega_{AB}$ ($s_{AB}=\pm 1$).
In this section we only consider the case (i).
The cases (i) and (ii) are equivalent in the massless limit
due to the SU(4) R-symmetry of the undeformed theory.
The corresponding configurations are interpreted as
the intersecting M2-branes spanning all the transverse coordinates~\cite{Fujimori:2010ec}.
We will discuss the case (ii) in the mass-deformed theory
in Appendix \ref{otherN=1}.
The BPS equations for the cases (iii) and (iv) are equivalent to
those of the ${\cal N}=2$ supersymmetries with conditions
$\omega_{13}=0$ and $\omega_{12}=0$, respectively.
We postpone the discussions on these phenomena to section \ref{halfint}.

\subsection {BPS equations and bound}\label{ss51}

When $\omega_{13}=\omega_{14}=0$, the BPS equations
are given by
\begin{alignat}{3}
& (D_{1}-isD_{2})Y^{a}=0,~~(a=1,2), \qquad &&
(D_{1}+isD_{2})Y^{p}=0,~~(p=3,4),
\nonumber\\
& D_{0}Y^{1}+is(\beta^{C1}_{\; C} -2\beta^{21}_{\; 2}-\mu Y^{1})=0,
\quad && D_{0}Y^{2}+is(\beta^{C2}_{\; C} -2\beta^{12}_{\; 1}-\mu
Y^{2})=0,
\nonumber\\
& D_{0}Y^{3}-is(\beta^{C3}_{\; C}-2\beta^{43}_{\; 4}+\mu Y^{3})=0,
\quad && D_{0}Y^{4}-is(\beta^{C4}_{\; C}-2\beta^{34}_{\; 3}+\mu
Y^{4})=0,
\nonumber\\
& \beta^{34}_{\; 1}=\beta^{34}_{\; 2}=\beta^{12}_{\;
3}=\beta^{12}_{\; 4}=0,\quad  (s=\pm  1).
\label{N1}
\end{alignat}
Note that in this case, all four scalar fields enter the equations
in a nontrivial way, i.e., satisfy the gauged Cauchy-Riemann equations
in contrast with $\mathcal{N}=3$ and $\mathcal{N}=2$ cases where only
one or two scalars have nontrivial profiles.

In accordance with \eqref{N1},
the energy expression for bosonic sector \eqref{Ene} can be reshuffled as
\begin{align}\label{ene71}
E =&\int d^2x \, {\rm tr} \,{\Bigg[} \sum_{a=1,2} \left|D_0 Y^{a}
          -is\left( \beta^{Aa}_{\;A} - 2\beta^{ba}_{\;b}
          - \mu Y^{a} \right) \right|^2
          +\sum_{p=3,4} \left|D_0 Y^{p}
          -is\left( \beta^{Ac}_{\;A} - 2\beta^{qp}_{\;q}
          + \mu Y^{p} \right) \right|^2
          \nonumber\\
          &\hspace{20mm}+\left.\sum_{a=1,2}\left|(D_1-isD_2)Y^a\right|^2
          +\sum_{p=3,4}\left|(D_1+isD_2)Y^p\right|^2
          \right. \nonumber\\
          &\hspace{20mm}+4\left(|\beta^{12}_{\;3}|^2+|\beta^{12}_{\;4}|^2
          +|\beta^{34}_{\;1}|^2+|\beta^{34}_{\;2}|^2\right){\Bigg]}\nonumber\\
&+is \,{\rm tr}\int d^2x\left[\epsilon_{ij}\partial_{i}(Y_a^\dag D_j
Y^a-Y_p^\dag D_j Y^p)\right]+s\mu\, {\rm tr}\int d^2x\,j^0,
\end{align}
where $a,b=1,2$ and $p,q=3,4$.
As we discussed previously, in the massless limit $\mu\to 0$, the energy is bounded
by the total derivative term in the fourth line of (\ref{ene71}).
In this section, however, we only consider the cases with mass deformation.
For any well behaved ${\cal N}=1$ BPS configurations with mass deformation,
the total derivative term does not contribute to the energy and then the energy is bounded by the U(1) charge \eqref{u1c},
\begin{equation}
E\geq \left|\mu Q_{{\rm U}(1)}\right| \label{EQ}.
\end{equation}
By the Gauss' law \eqref{Beq}, one can say that the energy is
bounded by the magnetic flux \eqref{mfu1},
\begin{equation}
E\geq \left|\frac{k\mu}{2\pi} \Phi_{{\rm U}(1)}\right| \label{EPh}.
\end{equation}
The bound in \eqref{EPh} is useful when we discuss topological
vortices carrying quantized magnetic flux, and the bound in
\eqref{EQ} is useful when we discuss nontopological $Q$-balls and
$Q$-vortices stabilized by conserved charge.

It is a straightforward matter to rewrite the spatial stress components of
energy-momentum tensor (\ref{stress}) as
\begin{align}\label{1Tij}
T_{ij}=&{\eta_{ij}}\,{\rm Re\, tr} \Big\{ \left[
(D_0 Y^a+ is(\beta^{Aa}_{\;A}-2\beta^{ba}_{\;b}-\mu Y^a))
(D_0 Y^a - is(\beta^{Aa}_{\;A}-2\beta^{ba}_{\;b}-\mu Y^a))^\dag \right.\notag\\
&\hspace{13mm} + \left. (1,2,\mu \leftrightarrow 3,4, -\mu) \right] \notag\\
&\hspace{13mm}-4\,(|\beta^{12}_{\;3}|^2+|\beta^{12}_{\;4}|^2+|
\beta^{34}_{\;1}|^2+|\beta^{34}_{\;2}|^2)\Big\} \notag\\
&+\frac{1}{2}\,{\rm Re\, tr} \left\{
[(D_i - is\epsilon_{ik} D_k)Y_A]^\dag(D_j + is\epsilon_{jl} D_l)Y^A
+(i \leftrightarrow j) \right\}.
\end{align}
Therefore, for any BPS soliton or anti-soliton configurations satisfying the
${\cal N}=1$ BPS equations \eqref{N1}, the spatial stress components of
energy-momentum tensor (\ref{1Tij}) vanish everywhere, as it should.

\subsection{BPS objects in the mass-deformed theory}\label{ss53}

The ${\cal N}=1$ BPS equations \eqref{N1} include four algebraic
constraints of which the number is much less than that of the
${\cal N}=2$ or ${\cal N}=3$ BPS equations.
Then even for the case of U(2)$\times$U(2) gauge group, it would not be
feasible to solve the BPS equations in the most general way.
As we did for the ${\cal N}=2$ case, here we will be content with simple cases.
Using ans\"atze based on the vacuum solutions \eqref{gvac},
we investigate possible BPS solutions from the cases of U(2)$\times$U(2) and U(3)$\times$U(3) gauge groups and then extend the results to the case of U($N$)$\times$U($N$) gauge group.

For the case of U(2)$\times$U(2) gauge group,
the vacuum configuration \eqref{gvac} suggests
\begin{align} \label{N1a}
Y^1=\sqrt{ \frac{k \mu}{2\pi} }\begin{pmatrix}
  0 & b\\ 0 &
  0
  \end{pmatrix},\qquad Y^2=\sqrt{ \frac{k \mu}{2\pi} }\begin{pmatrix}
  0 & 0\\ 0 &
  d
  \end{pmatrix},\qquad Y^3=Y^4=0.
\end{align}
up to a substitution $Y^{1,2}\leftrightarrow Y_{3,4}^\dag$. Then the
BPS equations \eqref{N1} are simplified to the following form:
\begin{equation}
(D_{1}- is D_{2})Y^{a} = 0,\quad
D_{0}Y^{a} - is (\beta^{ba}_{\;b} + \mu Y^{a}) = 0, \quad
a,b=1,2.
\label{n1bps}
\end{equation}
The resultant BPS equations are equivalent to those of ${\cal N}=2$
case in section \ref{sec4}. Therefore, there is no genuine ${\cal N}=1$
U(2)$\times$U(2) BPS solution within the ansatz \eqref{N1a} based on
the vacuum solution.
In fact, this is natural considering that all four scalar fields have to
be nontrivial for $\mathcal{N}=1$ solutions, as pointed out below \eqref{N1},
namely all four $Y^A$'s satisfy the gauged Cauchy-Riemann equations.

The lowest rank gauge group, for which all scalar fields are nonvanishing
within the ansatz based on the vacuum configurations, is U(3)$\times$U(3).
In this case we expect that there exist some $\mathcal{N}=1$ BPS solutions
with nontrivial configurations for all $Y^A$'s in this gauge group.
An interesting configuration can be obtained from the following ansatz based
on one of the vacuum solutions of the U(3)$\times$U(3) case,
\begin{align}\label{U3ans}
&Y^1 = \sqrt{ \frac{k \mu}{2\pi} }
      \begin{pmatrix} 0 & 0 & 0\\ 0 & 0 & a\\0 & 0 & 0\end{pmatrix},\quad
Y^2 = \sqrt{ \frac{k \mu}{2\pi} }
      \begin{pmatrix} 0 & 0 & 0\\ 0 & 0 & 0\\0 & 0 &
      b\end{pmatrix},\nonumber\\
&Y^3 = \sqrt{ \frac{k \mu}{2\pi} }
      \begin{pmatrix} d & 0 & 0\\ 0 & 0 & 0\\0 & 0 & 0\end{pmatrix},\quad
Y^4 = \sqrt{ \frac{k \mu}{2\pi} }
      \begin{pmatrix} 0 & e & 0\\ 0 & 0 & 0\\0 & 0 & 0\end{pmatrix},
\end{align}
where $a,b,d,e$ are arbitrary complex functions.
By the procedure similar to those in the previous sections we
obtain two pairs of coupled second order equations,
\begin{align}
&\partial\bar\partial \,{\rm
ln}\frac{|a|^2}{\displaystyle{\prod_{p=1}^{n_a}}|z-z^{(1)}_p|^{2}}
=\mu^2|b|^2(|a|^2-1),\qquad \partial\bar{\partial}\, {\rm{ln}}
\frac{ |b|^2}{ {\displaystyle{\prod_{q=1}^{n_b}}|z-z^{(2)}_q|^{2}}}=\mu
^2|a|^2(|b|^2-1), \label{eom11}\\
&\partial\bar{\partial}\, {\rm{ln}} \frac{ |d|^2}{
{\displaystyle{\prod_{r=1}^{n_d}}|z-z^{(3)}_r|^{2}}}=\mu
^2|e|^2(|d|^2-1), \qquad \partial\bar{\partial}\, {\rm{ln}} \frac{
|e|^2}{ {\displaystyle{\prod_{s=1}^{n_f}}|z-z^{(4)}_s|^{2}}}=\mu
^2|d|^2(|e|^2-1). \label{eom12}
\end{align}
Note that each pair is identical to
\eqref{eom2-5}--\eqref{eom2-6} in ${\cal N}=2$ case and hence the same
solutions.
In the present case, one pair of equations \eqref{eom11} lies on the
($Y^1$,$Y^2$)-plane while the other pair of equations \eqref{eom12} lies
on the ($Y^3$,$Y^4$)-plane.
The energy and angular momentum of this configuration are given by
\begin{align}
E=& k\mu\, (n_a+n_b+n_d+ n_e+\alpha_a+\alpha_b+\alpha_d +\alpha_e),
\nonumber\\
J=&k(\alpha_a\alpha_b-n_a n_b+\alpha_d\alpha_e-n_d n_e),
\end{align}
where $\alpha$ and $n$ are defined in \eqref{bdaa} and \eqref{flx}, respectively, and we calculated the angular momentum $J$ for the rotationally symmetric
case as before.

We can generalize the ansatz \eqref{U3ans} of the U(3)$\times$U(3) theory
to U($N$)$\times$U($N$) case as follows:
\begin{align}
Y^1&=\sqrt{ \frac{k \mu}{2\pi} }\left(\begin{array}{c}
\begin{array}{cccccc}{\bf 0}_{m_1\times (m_1+1)}&&&&&\\&\ddots&&&&\\
&&{\bf 0}_{m_I\times(m_I + 1)} &&&\\
&&& a_1{\mathcal M^\dag}_1^{(m_{I+1})}\!\!&&\\&&&&\!\!\ddots\!&\\
&&&&&\!\!a_K{\mathcal M^\dag}_1^{(m_{I+K})}\end{array}\\
\end{array}\right),\nonumber\\
Y^2&=\sqrt{ \frac{k \mu}{2\pi} }\left(\begin{array}{c}
\begin{array}{cccccc}{\bf 0}_{m_1\times (m_1+1)}&&&&&\\&\ddots&&&&\\
&&{\bf 0}_{m_I\times(m_I + 1)} &&&\\
&&& b_1{\mathcal M^\dag}_2^{(m_{I+1})}\!\!&&\\&&&&\!\!\ddots\!&\\
&&&&&\!\!b_K{\mathcal M^\dag}_2^{(m_{I+K})}\end{array}\\
\end{array}\right),\nonumber
\end{align}
\vskip -0.5cm
\begin{align}
Y^3&=\sqrt{ \frac{k \mu}{2\pi} } \left(\begin{array}{c}
\begin{array}{cccccc}d_1\mathcal{M}_1^{(m_1)}\!\!&&&&&\\&\!\!\ddots\!&&&&\\
&&\!\!d_I\mathcal{M}_1^{(m_I)}&&& \\ &&& {\bf 0}_{(m_{I+1}+1)\times m_{I+1}}
&&\\&&&&\ddots&\\&&&&&{\bf 0}_{(m_{I+K}+1)\times m_{I+K}}\end{array}\\
\end{array}\right),\nonumber\\
~~~~~~~~Y^4&=\sqrt{ \frac{k \mu}{2\pi} } \left(\begin{array}{c}
\begin{array}{cccccc}e_1\mathcal{M}_2^{(m_1)}\!\!&&&&&\\&\!\!\ddots\!&&&&\\
&&\!\!e_I\mathcal{M}_2^{(m_I)}&&& \\ &&& {\bf 0}_{(m_{I+1}+1)\times m_{I+1}}
&&\\&&&&\ddots&\\&&&&&{\bf 0}_{(m_{I+K}+1)\times m_{I+K}}\end{array}\\
\end{array}\right),\label{Nans2}
\end{align}
where $a_k,b_k$ ($k=1,\cdots,K)$ and  $d_i,e_i$ ($i=1,\cdots,I)$ are arbitrary complex functions.
Then we have ($K+I$)-pairs of coupled differential equations
\begin{align}
&\partial\bar\partial \,{\rm
ln}\frac{|a_k|^2}{\displaystyle{\prod_{p_k=1}^{n_{a,k}}}|z-z^{(1)}_{p_k}|^{2}}
=\mu^2|b_k|^2(|a_k|^2-1),\qquad \partial\bar{\partial}\, {\rm{ln}}
\frac{ |b_k|^2}{ {\displaystyle{\prod_{q_k=1}^{n_{b,k}}}|z-z^{(2)}_{q_k}|^{2}}}=\mu
^2|a_k|^2(|b_k|^2-1), \nonumber\\
&\partial\bar{\partial}\, {\rm{ln}} \frac{ |d_i|^2}{
{\displaystyle{\prod_{r_i=1}^{n_{d,i}}}|z-z^{(3)}_{r_i}|^{2}}}=\mu
^2|e_i|^2(|d_i|^2-1), \qquad \partial\bar{\partial}\, {\rm{ln}} \frac{
|e_i|^2}{ {\displaystyle{\prod_{s_i=1}^{n_{e,i}}}|z-z^{(4)}_{s_i}|^{2}}}=\mu
^2|d_i|^2(|e_i|^2-1). \label{eomN}
\end{align}
Here, the $K$-pairs of equations in the first line of \eqref{eomN} live on the
($Y^1$,$Y^2$)-plane and the $I$-pairs of equations in the second line of \eqref{eomN} live on the ($Y^3$,$Y^4$)-plane. The energy and angular momentum of this configuration are
\begin{align}
E=&k\mu{\displaystyle\sum_k^K}\frac{m_k(m_k+1)}{2}  (n_{a,\,k}+n_{b,\,k}+\alpha_{a,\,k}+\alpha_{b,\,k})
+k\mu{\displaystyle\sum_i^I}\frac{m_i(m_i+1)}{2} (n_{d,\,i}+n_{e,\,i}+\alpha_{d,\,i}+\alpha_{e,\,i}),\nonumber\\
J=&k{\displaystyle\sum_k^K}\frac{m_k(m_k+1)}{2}(\alpha_{a,\,k}\alpha_{b,\,k}-n_{a,\,k} n_{b,\,k})+k{\displaystyle\sum_i^I}\frac{m_i(m_i+1)}{2} (\alpha_{d,\,i}\alpha_{e,\,i}
-n_{d,\,i} n_{e,\,i}),\label{EJN1}
\end{align}
where for the calculation of angular momentum we again considered
rotationally symmetric configurations only.

When we fix $d_k=e_k=1$, the ansatz \eqref{Nans2} and the corresponding physical quantities
are equivalent to case of ${\cal N}=2$ in section~\ref{sec4} and hence
the supersymmetry of the BPS solutions in this case is enhanced to ${\cal N}=2$.

\section{Absence of Vortex-type Objects with ${\cal N}=\frac52$, $\frac32$, $\frac12$ Supersymmetries}\label{halfint}

In this section, we discuss the BPS nature of half integer supersymmetries,
which are obtained by imposing the supersymmetric conditions  $\gamma^1\omega_{AB} =\pm\omega_{AB}^*$,
in addition to the conditions of integer supersymmetries.
These additional conditions deform the gauged Cauchy-Riemann equations in the BPS equations
of integer supersymmetries. From the algebraic relations of BPS equations and the vanishing
$T_{ij}$ condition discussed in section \ref{ss31}, we will argue that the spectrum of BPS solitons of half-integer supersymmetries are equivalent to those of integer supersymmetries.

\setcounter{equation}{0}
\subsection{$\cal N$=$\frac{5}{2}$ supersymmetry }\label{sec6}

As we see in Fig.1, there are two ways to obtain ${\cal N}=\frac52$
BPS equations.
In addition to $\gamma^0\omega_{AB} = is_{AB} \omega_{AB}$ which is
necessary to get vortex-type equations, we can
impose one of the following conditions

(i) $\gamma^1\omega_{12} = \omega_{34}$,

(ii) $\gamma^1\omega_{13} = \omega_{24}$.

\noindent
One of these would kill one real degree of $\omega_{12}$ or $\omega_{13}$,
leaving five real independent supersymmetric parameters.
Here, we will consider
the case (i). The same argument can be applied to the case (ii) as well.

In the presence of the additional condition (i), the BPS equations
are the same as
those in \eqref{hpeq} except the Cauchy-Riemann equation for
$Y^2$, which is modified as
\begin{align}
(D_{1}-isD_{2})Y^{2}=0
\quad \longrightarrow \quad (D_{1}-isD_{2})Y^{2}=2\beta^{34}_{\; 1}, \label{N512}
\end{align}
and a constraint $\beta^{34}_{\;1}=0$, which is no longer zero.

It should be noted however that the resulting BPS equations do not
necessarily have nontrivial solutions with the
expected number of supersymmetries. In other words, all the solutions
may have enhanced supersymmetries and hence the BPS equations
are actually equivalent to those of higher symmetries.
This is indeed the case with $\mathcal{N}=\frac52$ equations which are
actually equivalent to $\mathcal{N}=3$ equations. To see this,
we multiply $(\beta^{34}_{\;1})^{\dag}$ to the deformed BPS equation
\eqref{N512} and take trace,
\begin{align}
\mathrm{tr}|\beta^{34}_{\;1}|^{2} &= \frac{1}{2}\mathrm{tr}
\bigl[(\beta^{34}_{\;1})^{\dagger}(D_{1}-isD_{2})Y^{2}\bigr]
\notag\\
&= \frac{1}{2}(\partial_{1}-is\partial_{2})
\mathrm{tr}\bigl[(\beta^{34}_{\;1})^{\dagger}Y^{2}\bigr]
\notag\\
&\quad +\frac{1}{2}\mathrm{tr}\bigl[
\beta^{12}_{\;4}(D_{1}-isD_{2})Y^{\dagger}_{3}
+(\beta^{34}_{\;2})^{\dagger}(D_{1}-isD_{2})Y^{1}
-\beta^{12}_{\;3}(D_{1}-isD_{2})Y^{\dagger}_{4} \bigr]. \label{512E}
\end{align}
Note that each term in the second trace vanishes due to the BPS
equations for $Y^1, Y^3, Y^4$ in \eqref{hpeq} and there remains only the
first term. But $\mathrm{tr}\bigl[(\beta^{34}_{\;1})^{\dagger}Y^{2}\bigr]
=-\mathrm{tr}\bigl[(\beta^{34}_{\;2})^{\dagger}Y^{1}\bigr]$, which
again is zero thanks to the constraint $\beta^{34}_{\;2}=0$. Therefore we
reobtain the the missing constraint,
\begin{gather}
\beta^{34}_{\;1}=0, \label{B314}
\end{gather}
which results in $\mathcal{N}=3$ equations.

At this point, it is worth examining the BPS nature of ${\cal N}=\frac52$
from the point of view of the stress components of energy-momentum tensor,
$T_{ij}$. In section \ref{ss31}, we discussed the relation between force
and stress components of energy-momentum tensor.
From \eqref{forc}, we read vanishing $T_{ij}$ as a
sufficient condition for noninteracting BPS solitons.
The terms in the spatial stress components of
energy-momentum tensor \eqref{eneT} can be reshuffled as
\begin{align}
T_{ij}= \frac13 {\rm Re\; tr} & \left\{  \eta_{ij} \left[ \left(
         \delta_{A}^{[ B} D_0 Y^{C]} +is_{ B C} ( \beta^{ B C}_{\;A}
         + \delta^{[B}_{A} \beta^{C] D}_{\;D} + \mu M_{ A}^{\;[B} Y^{C]} )
         \right)^\dag \right. \right. \notag\\
&\qquad \left. \times \left(
         \delta_{A}^{[B} D_0 Y^{C]} -is_{ BC} ( \beta^{BC}_{\;A}
        + \delta^{[B}_{A} \beta^{C] D}_{\;D} + \mu M_{A}^{\;[B} Y^{C]} )
         \right) \right] \notag \\
&    + ((D_i + is\epsilon_{ik} D_k) Y^A )^\dag (D_j - is\epsilon_{jl} D_l) Y^A
      + ((D_i - is\epsilon_{ik} D_k) Y^A )^\dag (D_j + is\epsilon_{jl} D_l) Y^A  \notag\\
& -(is)^{i+j}\left[
  ( (D_1 -is D_2)Y^a + 2 \epsilon^{ab} \beta^{34}_{\;b})^\dag
  (D_1 + isD_2)Y^a \right. \notag \\
&\hspace{17mm} +( (D_1 + isD_2)Y^p + 2 \epsilon^{pq} \beta^{12}_{\;q} )
  ( (D_1 -is D_2)Y^p )^\dag  \notag \\
&\hspace{17mm} +\left. 2 \epsilon^{ab} (\beta^{34}_{\;a})^\dag (D_1+is D_2)Y^b
 + 2 \epsilon^{pq} \beta^{12}_{\;p} ( (D_1-is D_2)Y^q )^\dag \right] \notag \\
& -4  \eta_{ij} |\beta^{34}_{1}|^2 \bigg\},  \label{Str52}
\end{align}
where $a,b=1,2$ and $p,q=3,4$.
On imposing the original form of $\mathcal{N}=\frac52$ BPS equations,
all terms except the last term vanish and we are left with
\begin{align}
T_{ij}=-\frac43\eta_{ij} \rm{tr} |\beta^{34}_{\;1}|^2.
\end{align}
As seen above, however, the consistency of the equations requires \eqref{B314}
and it precisely corresponds to the condition that the stress tensor
should vanish to have noninteracting BPS solitons. We will see that
this kind of structure reappears in other cases with half-integer supersymmetry.
In fact, it turns out that the manipulation of the stress tensor is
quite a useful tool to obtain consistency conditions of BPS equations.

\subsection{$\cal N$=$\frac 32$ supersymmetry}\label{sec7}

There are four ways to obtain ${\cal N}=\frac32$ BPS equations:

(i) $\omega_{12} = 0$, $\gamma^1\omega_{13} = \omega_{24}$,

(ii) $\omega_{13} = 0$, $\gamma^1\omega_{12} = \omega_{34}$,

(iii) $\omega_{13} = 0$, $\gamma^1\omega_{14} = \omega_{23}$,

(iv) $\gamma^1\omega_{12} = \omega_{34}$,
$\gamma^1\omega_{13} = \omega_{42}$, $\gamma^1\omega_{14}= \omega_{23}$

\noindent
in addition to the condition $\gamma^{0}\omega_{AB} = is_{AB}\omega_{AB}$ ($s_{AB}=\pm 1$).
We will treat the case (i) and (iv). The cases (ii) and (iii) are similar to the case (i).

\subsubsection {$\omega_{12} = 0$, $\gamma^1\omega_{13} = \omega_{24}$ case}\label{ss71}

In this case the BPS equations are the same as those of the
$\mathcal{N}=2$ case \eqref{N22} except that the gauged Cauchy-Riemann
equations in the second line of \eqref{N22} are changed to
\begin{align}
&(D_{1}-isD_{2})Y^{3}=0\quad \longrightarrow \quad
(D_{1}-isD_{2})Y^{3}=-2\beta^{24}_{\;1},\label{N323}\\
&(D_{1}-isD_{2})Y^{4}=0 \quad \longrightarrow \quad (D_{1}-isD_{2})Y^{4}=-2\beta^{13}_{\;2},\label{N324}
\end{align}
and two algebraic constraints $\beta^{24}_{\;1}=0$ and
$\beta^{13}_{\;2}=0$ disappear.
As we did in $\mathcal{N}=\frac52$ case, we rewrite a positive semi-definite
quantity using the deformed equations \eqref{N323} and \eqref{N324},
\begin{align}
\mathrm{tr}\bigl[|\beta^{24}_{\;1}|^{2}+|\beta^{13}_{\;2}|^{2}\bigr]
&= -\frac{1}{2}\mathrm{tr}
\bigl[(\beta^{24}_{\;1})^{\dagger}(D_{1}-isD_{2})Y^{3}
+\beta^{13}_{\;2}(D_{1}+isD_{2})Y_{4}^{\dagger}\bigr]
\notag\\
&= -\frac{1}{2}(\partial_{1}-is\partial_{2})
\mathrm{tr}\bigl[(\beta^{24}_{\;1})^{\dagger}Y^{3}\bigr]
\notag\\
&\quad\ {} -\frac{1}{2}\mathrm{tr}\bigl[
(\beta^{24}_{\;3})^{\dagger}(D_{1}-isD_{2})Y^{1}
+\beta^{13}_{\;4}\bigl((D_{1}+isD_{2})Y^{2}\bigr)^{\dagger} \bigr].
\label{32E}
\end{align}
The second trace of \eqref{32E} vanish
due to the gauged Cauchy-Riemann equations in the first line of
\eqref{N22}. Moreover, since
$\mathrm{tr}\bigl[(\beta^{24}_{\;1})^{\dagger}Y^{3}\bigr]
=-\mathrm{tr}\bigl[(\beta^{24}_{\;3})^{\dagger}Y^{1}\bigr]$ and
$\beta^{24}_{\;3}=0$ still holds, the right hand side vanishes identically.
Thus two missing constraints are regained,
\begin{gather}
\beta^{24}_{\;1}=\beta^{13}_{\;2}=0. \label{2413}
\end{gather}
Substituting \eqref{2413} into \eqref{N323}--\eqref{N324} the set
of BPS equations for $\mathcal{N}=\frac{3}{2}$ supersymmetry
coincides with that for $\mathcal{N}=2$ supersymmetry. Therefore there
is no solution with genuine $\mathcal{N}=\frac{3}{2}$ supersymmetry.

As before, this can be seen by considering the stress tensor.
Inserting the $\mathcal{N}=\frac{3}{2}$ BPS equations
into the expression of $T_{ij}$, we obtain
\begin{align}
T_{ij}=-2\eta_{ij}\rm{tr}(|{\beta^{24}_{\;1}}|^2+| {\beta^{13}_{\;2}}|^2),
\end{align}
which vanishes.

\subsubsection{$\gamma^1\omega_{12} = \omega_{34}$,
$\gamma^1\omega_{13} = \omega_{42}$, $\gamma^1\omega_{14}= \omega_{23}$ case}%
\label{SA3}

The supersymmetric condition for this case has an SO(3) symmetry with
respect to the indices 2, 3 and 4 and it is reflected on the BPS
equations:
\begin{align}
&(D_{1}-isD_{2})Y^{1}=2\beta^{42}_{\;3} = 2\beta^{34}_{\;2}=2\beta^{23}_{\;4} \nonumber\\
&(D_{1}-isD_{2})Y^{2}=-2\beta^{34}_{\;1},\qquad
\hspace{7mm}(D_{1}+isD_{2})Y^{2}=-2\beta^{13}_{\;4}=2\beta^{14}_{\;3},\nonumber\\
&(D_{1}-isD_{2})Y^{3}=-2\beta^{42}_{\;1},\qquad
\hspace{4mm}(D_{1}+isD_{2})Y^{3}=-2\beta^{14}_{\;2}=2\beta^{12}_{\;4},\nonumber\\
&(D_{1}-isD_{2})Y^{4}=-2\beta^{23}_{\;1},\qquad
\hspace{4mm}(D_{1}+isD_{2})Y^{4}=-2\beta^{12}_{\;3}=2\beta^{13}_{\;2},\nonumber\\
&D_0 Y^1 +is (\beta^{21}_{\;2} + \mu Y^1)  = 0, \qquad
D_0 Y^2 -is (\beta^{12}_{\;1} + \mu Y^2)  = 0, \nonumber \\
&D_0 Y^3 -is \beta^{13}_{\;1} = 0, \qquad
\hspace{15mm}D_0 Y^4 -is \beta^{14}_{\;1} = 0, \nonumber \\
&\beta^{31}_{\;3} = \beta^{41}_{\;4} = \beta^{21}_{\;2} + \mu Y^1,
\qquad \hspace{6mm}\beta^{43}_{\;4} = \mu Y^3, \qquad
\beta^{34}_{\;3} = \mu Y^4, \nonumber \\
&\beta^{32}_{\;3} = \beta^{42}_{\;4} = \beta^{23}_{\;2} =
\beta^{24}_{\;2} = 0.\label{N322}
\end{align}
When we set the $\beta^{BC}_{\;A}=0$ $(A\neq B\neq C\neq A)$,
the BPS equations \eqref{N322} are the same as those of
${\cal N}=3$ case in \eqref{hpeq}.

One may wonder whether there is any supersymmetry enhancement due to
consistency between the equations as in the previous subsections. The
answer turns out to be positive but the argument is rather subtle in this case.
We first calculate the stress tensor using \eqref{N322},
\begin{equation} \label{N3231}
T_{ij} = - \frac23 \eta_{ij}{\rm{tr}} \sum_{A\neq B\neq C \neq A}
          |\beta^{AB}_{\;C}|^2.
\end{equation}
Note that, as in the previous cases, the summation consists of positive
semi-definite terms which
are actually the missing constraints not existing in \eqref{N322}.
Then one may suspect that the missing constraints should be obtained
from \eqref{N322} by similar manipulations to \eqref{32E}. After inserting
\eqref{N322} into \eqref{N3231}, we find
\begin{equation} \label{N3241}
T_{ij} = - \frac13 \eta_{ij}{\rm{tr}} \epsilon_{ijk}
         (\partial_1 - is\partial_2)[(\beta_i^{ij})^\dag Y^1],
\end{equation}
where $i,j,k=2,3,4$. From this expression, we cannot directly conclude
$T_{ij}=0$ because none of the terms can be put to zero by itself. Note,
however, that the terms inside the total derivative vanish for
vacuum configurations, see \eqref{ve3}. Integrating over the whole space,
it then should go to zero for finite energy configurations, i.e.,
\begin{equation}
\int d^2x T_{ij} = 0.
\end{equation}
From \eqref{N3231}, it is now obvious that each positive semi-definite term
should vanish: $\beta^{BC}_{\;A}=0$ $(A\neq B\neq C\neq A)$. This concludes
the argument that the case (iv) with $\mathcal{N}=3/2$ supersymmetry
is actually enhanced to $\mathcal{N}=3$ case.

\setcounter{equation}{0}
\subsection{$\cal N$=$\frac{1}{2}$ supersymmetry}\label{sec8}

There are two ways to obtain ${\cal N}=\frac12$ BPS equations:

(i) $\omega_{12} = 0$, $\omega_{13} = 0$, $\gamma^1\omega_{14} = \omega_{23}$,

(ii) $\omega_{13} = 0$, $\omega_{14} = 0$, $\gamma^1\omega_{12} = \omega_{34}$,

\noindent
under the condition $\gamma^{0}\omega_{AB} = is_{AB}\omega_{AB}$
($s_{AB}=\pm 1$). Here we will consider only the case (i). The case (ii) is
similar to case (i).

The BPS equations of the case (i) are almost the same as those
of ${\cal N}=1$ case; the only change is the deformation of gauged
Cauchy-Riemann equations in the first line of \eqref{N1}
\begin{align}
&(D_{1}-isD_{2})Y^{a}=0\quad \longrightarrow \quad
(D_{1}-isD_{2})Y^{a}=2 \epsilon^{ab}\beta^{34}_{\; b}, \quad (a,b=1,2),\\
&(D_{1}+isD_{2})Y^{p}=0\quad \longrightarrow \quad
(D_{1}+isD_{2})Y^{p}=2 \epsilon^{pq}\beta^{12}_{\; q}, \quad (p,q=3,4)\label{Bog12}
\end{align}
with nonvanishing $\beta^{34}_{\; b}$ and $\beta^{12}_{\; q}$.
Like the case discussed in section \ref{SA3}, we cannot fix
$\beta^{34}_{\; b}$ and $\beta^{12}_{\; q}$ to zero
by algebraic manipulations of the BPS equations.
In order to figure out the BPS nature of ${\cal N}=\frac12$ object
we calculate $T_{ij}$.  Applying the BPS equations into the
$T_{ij}$, we have
\begin{align}
T_{ij}&=-4\eta_{ij}{\rm{tr}}\left(|\beta^{12}_{\;3}|^2 +|\beta^{12}_{\;4}|^2
          +|\beta^{34}_{\;1}|^2+|\beta^{34}_{\;2}|^2\right) \notag \\
      &=2\eta_{ij}{\rm tr} (\partial_1 - i \partial_2 )
        [ (\beta^{34}_{\;1})^\dag Y^2 ],
\end{align}
which is again a summation of positive semi-definite terms and, at the
same time, a total derivative of a term vanishing for vacuum configurations.
Therefore we recover the missing constraints, $\beta^{34}_{\;b}=0$
and $\beta^{12}_{\;q}=0$. Then the supersymmetry is actually enhanced to
${\cal N}=1$.

\setcounter{equation}{0}
\section{Conclusion}\label{sec9}

We investigated the vortex-type BPS equations with various supersymmetries
in the ABJM theory without or with mass-deformation.
For a given number of supersymmetry, we classified distinguishable
BPS conditions, and then obtained the BPS equations
and the energy bound.
As a nontrivial consistency check of the BPS equations, we investigated
the stress components of energy-momentum tensor and showed that it vanishes.
Then, we set a special type of ansatz which solves constraints of the BPS
equations, based on the discrete vacua of the mass-deformed ABJM theory.
Using these ans$\ddot{{\rm a}}$tze for U($N)\times{\rm U}(N)$ gauge group,
we obtained several types of BPS vortex equations with finite energy.

For the undeformed ABJM theory we obtained ${\cal N}=2$ BPS equations.
After solving all the constraint equations of the BPS equations
for U(2)$\times$U(2) gauge group, we showed that the resulting equations
are reduced to the Liouville-type or Sinh-Gordon-type vortex equations
in special limits.

For the mass-deformed theory with U($N)\times{\rm U}(N)$ gauge group,
we obtained special types of ${\cal N}=3,2,1$ BPS configurations.
In constructing these configurations, we used ans$\ddot{{\rm a}}$tze
based on the vacuum solutions to solve the complicated constraint equations
and to obtain finite energy configurations.
Our BPS vortex equations are summarized as follows:
\begin{center}
\begin{tabular}{|c||c|c|}
\hline
& gamma matrix projection & vortex-type equation \\
\hline
${\cal N}$=$3$ &\small $\gamma^0\omega_{AB}$=$\pm i \omega_{AB}$ & $K$-MH on $Y^1$, vacua along $(Y^2, Y^3, Y^4)$\\
\hline
${\cal N}$=$2$ & \small$\gamma^0\omega_{AB}$=$\pm i \omega_{AB}$, $\omega_{12}=0$ &
$K$-pairs of CDE on $(Y^1,Y^2)$, vacua along $(Y^3, Y^4)$\\
\hline
${\cal N}$=$2$ &\small $\gamma^0\omega_{AB}$=$\pm i \omega_{AB}$, $\omega_{13}=0$ &
$K$-MH on $Y^1$, $I$-MH on $Y^3$, vacua along $(Y^2, Y^4)$\\
\hline
${\cal N}$=$1$ & \small $\gamma^0\omega_{AB}$=$\pm i \omega_{AB}$,  $\omega_{13}$=$\omega_{14}$=$0$ &
$K$-pairs of CDE on $(Y^1,Y^2)$, $I$-pairs of CDE on $(Y^3,Y^4)$ \\
\hline
${\cal N}$=$1$ & \small $\gamma^0\omega_{AB}$=$\pm i \omega_{AB}$, $\omega_{12}$=$\omega_{13}$=$0$ &
$K$-pairs of CDE on $(Y^1,Y^2)$, $I$-pairs of CDE on $(Y^3,Y^4)$ \\
\hline
\end{tabular}
\end{center}
where MH denotes the vortex equations in Maxwell-Higgs theory and CDE the coupled second-order differential equations discussed in section 4, and $K$ and $I$ indicate the numbers of nonvanishing
blocks of vacuum solutions on $(Y^1,Y^2)$- and $(Y^3,Y^4)$-planes, respectively. The two ${\cal N}=1$ cases are actually equivalent. See
Appendix \ref{otherN=1} for the details.

In section 6, we also analyzed the cases of half-integer supersymmetries.
With the help of the stress tensor $T_{ij}$, we showed that the supersymmetries
are actually enhanced to integer ones. In other words,
the BPS equations with $\mathcal{N}=\frac52$, $\mathcal{N}=\frac32$, and
$\mathcal{N}=\frac12$ supersymmetries are respectively equivalent to those
of $\mathcal{N}=3$, $\mathcal{N}=2$ or 3 (depending on the supersymmetry
conditions), and $\mathcal{N}=1$ supersymmetries.

The BPS configurations of the ${\cal N}=3,2,1$ BPS vortex equations in
the undeformed ABJM theory were interpreted as intersecting M2-branes spanning
one, two, and four complex coordinates in transverse directions,
respectively~\cite{Fujimori:2010ec}. However, the brane interpretation of
the BPS vortex equations in the mass-deformed ABJM theory in M-theory
$(k\ll N^{\frac15})$ is unclear up to now, though the configuration of
the ${\cal N}=3$ vortex equations in the Maxwell-Higgs
theory obtained in section 3 can be identified with D0-branes in type IIA
string theory ($N^{\frac15}\ll k\ll N$)~\cite{Auzzi:2009es}.
In this paper, we obtained some pairs of coupled differential equation, which
can be reduced to the vortex equations in Maxwell-Higgs
theory or Chern-Simons matter theories, in
special limits of the ${\cal N}=2,1$ vortex-type BPS equations.
These pairs of coupled differential equation reflect the complicated vacuum
structure~\cite{Gomis:2008vc,Kim:2010mr} of the mass-deformed ABJM theory.
It would be interesting if we can identify the corresponding configurations
for the coupled differential equation in dual gravity
limit~\cite{Bena:2004jw,Lin:2004nb,Cheon:2011gv}.

\section*{Acknowledgements}
This work has been supported by the National Research Foundation of
Korea(NRF) grant funded by the Korea government(MEST) 
(No. 2011-0011660) (Y.K.),
by the World Class University
grant R32-2008-000-10130-0 and the NRF grant funded by the
Korea government(MEST) (No. 2012-045385) (C.K.),
by the National Research Foundation of
Korea(NRF) grant funded by the Korea government(MEST) 
(No. 2011-0009972) (O.K.),
and by National Science Council and National Center for Theoretical Sciences,
Taiwan, R.O.C. (H.N.).

\appendix

\setcounter{equation}{0}
\section{Vortex-type Objects with Other ${\cal N}=1,2$ Supersymmetries}

We listed the various supersymmetric cases in Fig.1
by imposing different supersymmetric conditions
to the supersymmetric parameters $\omega^{AB}$.
We did not treat all of those cases in the body of this paper.
In this appendix, we briefly discuss their possible
soliton objects.

\subsection{$\cal N$=2  supersymmetry ($\omega_{13} =0$)}

{\bf$\bullet$ {BPS equations and BPS bound :}} In section \ref{sec4}, we obtained the BPS solutions for $\omega_{12} =0$ case. Here we will treat the $\omega_{13}=0$ case (or equivalently $\omega_{14}=0$ case which is connected by field redefinition, $Y^3\leftrightarrow Y^4$). Then the BPS equations of this case are given by
\begin{alignat}{3}
& (D_{1}- is D_{2})Y^{1}=0,  && (D_{1}+is D_{2})Y^{3}=0,
\nonumber\\
& D_{1}Y^{p}=D_{2}Y^{p}=0,~~(p=2,4), &&
\nonumber\\
& D_{0}Y^{1}+is\beta^{31}_{\; 3}=0,  &&
D_{0}Y^{2}-is(\beta^{12}_{\; 1}-\beta^{32}_{\; 3}+\mu Y^{2})=0,
\nonumber\\
& D_{0}Y^{3}-is\beta^{13}_{\; 1} =0,
&&D_{0}Y^{4}-is(\beta^{14}_{\; 1}-\beta^{34}_{\; 3}+\mu Y^{4})=0,
\nonumber\\
& \beta^{42}_{\; 4}=\beta^{24}_{\; 2}=0, &&\beta^{21}_{\;
2}-\beta^{41}_{\; 4} =-\mu Y^{1}, \qquad \beta^{23}_{\;
2}-\beta^{43}_{\; 4}=-\mu Y^{3},
\nonumber\\
& \beta^{23}_{\; 1}=\beta^{34}_{\; 1}=\beta^{34}_{\;
2}=\beta^{14}_{\; 2} =\beta^{12}_{\; 3} = \beta^{14}_{\; 3}=
\beta^{12}_{\; 4} &&=\beta^{23}_{\; 4}=0. \label{N23}
\end{alignat}

By using the BPS equations \eqref{N23} we
can show that the energy is bounded by R-charge,
\begin{align}
E \geq\mu R_{24}, \label{eneN22}
\end{align}
where
\begin{align}
R_{24} = \int d^2x\,{\rm tr}\, J_{24}^0 =\int d^2x\,{\rm
tr}\left[i(Y^2 D_0Y_2^\dagger - D_0 Y^2 Y_2^\dagger)
         + i(Y^4 D_0 Y_4^\dagger - D_0 Y^4 Y_4^\dagger)\right].\label{r24}
\end{align}
\noindent
{\bf $\bullet$ {BPS objects :}} As we did in previous sections, we consider special types of solutions,
based the vacuum solutions of the mass-deformed ABJM theory.
From the shapes of the gauged Cauchy-Riemanns equations in \eqref{N23}
one can naturally consider an ansatz, which is constructed from \eqref{Nans2}
by setting $b_k=1$ ($k=1,...,K$) and $e_i=1$ $(i=1,...,I)$.
From \eqref{eomN} we can easily read the corresponding second order differential equations
for scalar fields as
\begin{align}\label{ON2}
\partial\bar\partial \,{\rm
ln}\frac{|a_k|^2}{\displaystyle{\prod_{p_k=1}^{n_{a,k}}}|z-z_{p_k}|^{2}}
=\mu^2(|a_k|^2-1),\qquad
\partial\bar{\partial}\, {\rm{ln}} \frac{ |d_i|^2}{
{\displaystyle{\prod_{r_i=1}^{n_{d,i}}}|z-z_{r_i}|^{2}}}=\mu
^2(|d_i|^2-1).
\end{align}
The energy of this configuration is given by
\begin{align}
E=&k\mu{\displaystyle\sum_k^K}\frac{m_k(m_k+1) n_{a,\,k}}{2}
+k\mu{\displaystyle\sum_i^I}\frac{m_i(m_i+1)n_{d,\,i}}{2}.
\end{align}
The scalar equations in \eqref{ON2} represent two sets of Maxwell-Higgs type vortex equations
living on $(Y^1,Y^2)$- and $(Y^3,Y^4)$-planes.
Each set of the equations appears in the ${\cal N}=3$
BPS configurations. Since there exist only Maxwell-Higgs type of vortex equations,
the angular momentum of this configuration vanishes.

\subsection{${\cal N}=1$ supersymmetry ($\omega_{12}=\omega_{13}=0$)}\label{otherN=1}

{\bf $\bullet$ {BPS equations and BPS bound :}} If we choose $\omega_{12}=\omega_{13}=0$, the BPS
equations are
\begin{alignat}{3}
& (D_{1}-isD_{2})Y^{a}=0,~~(a=1,4), \qquad &&
(D_{1}+isD_{2})Y^{p}=0,~~(p=2,3),
\nonumber\\
& D_{0}Y^{1}+is(\beta^{C1}_{\; C} -2\beta^{41}_{\; 4}+\mu Y^{1})=0,
\quad && D_{0}Y^{2}-is(\beta^{C2}_{\; C} -2\beta^{32}_{\; 3}+\mu
Y^{2})=0,
\nonumber\\
& D_{0}Y^{3}-is(\beta^{C3}_{\; C}-2\beta^{23}_{\; 2}-\mu Y^{3})=0,
\quad && D_{0}Y^{4}+is(\beta^{C4}_{\; C}-2\beta^{14}_{\; 1}-\mu
Y^{4})=0,
\nonumber\\
& \beta^{23}_{\; 1}=\beta^{14}_{\; 2}=\beta^{14}_{\;
3}=\beta^{23}_{\; 4}=0. && \label{N12}
\end{alignat}
Then we can write the energy bound as
\begin{align}\label{eneB2}
E \geq \mu (R_{12}+R_{34}),
\end{align}
where $R_{12}+R_{34}$ is
\begin{align}
R_{12}+R_{34}=\int d^2x\,{\rm tr}\, (J_{12}^0+J_{34}^0) =\int d^2x\,{\rm
tr}&\left[i(Y^1 D_0Y_1^\dagger - D_0 Y^1 Y_1^\dagger)-i(Y^2 D_0Y_2^\dagger - D_0 Y^2 Y_2^\dagger)
        \right.\nonumber\\
         &\,\,\left.+i(Y^3 D_0 Y_3^\dagger - D_0 Y^3 Y_3^\dagger)- i(Y^4 D_0 Y_4^\dagger - D_0 Y^4 Y_4^\dagger)\right].\label{r1234}
\end{align}
\noindent
{\bf $\bullet$ {BPS objects :}} Note that if we change the relative signatures between $D_1 Y^{2,4}$ and $D_2 Y^{2,4}$ in the Cauchy-Riemann type equations in the first line of \eqref{N12}, they are the same as those of
$\cal N$=1 case with $\omega_{13}=\omega_{14}$=0 (see \eqref{N1}). Therefore the resulting
second order differential equations of this case are equivalent to the $\cal N$=1 case with $\omega_{13}=\omega_{14}$=0
up to substitutions $b_k\leftrightarrow b_k^\ast$ and $e_i\leftrightarrow e_i^\ast$ in the ansatz \eqref{Nans2}. As a result, in this case we have the same BPS soliton solutions
as those of ${\cal N}=1$ case in section \ref{sec5}.

\end{document}